\documentclass[prd,superscriptaddress,amsfonts,amssymb,amsmath,showpacs,onecolumn]{revtex4-2}
\usepackage{bm}
\usepackage{amsfonts}
\usepackage{latexsym}
\usepackage[latin1]{inputenc}
\usepackage{graphicx}
\usepackage{amsmath}
\usepackage{palatino}
\usepackage{mathpazo}
\usepackage{textcomp}
\usepackage{mathtools}
\usepackage{float}
\usepackage{booktabs}
\usepackage{dcolumn}
\usepackage{hyperref}
\hypersetup{colorlinks,citecolor=blue}
\usepackage{amsmath}
\usepackage{xcolor}
\usepackage{orcidlink}
\usepackage{caption}
\usepackage{subcaption}
\usepackage{commath}
\def\jnl@style{\it}
\def\aaref@jnl#1{{\jnl@style#1}}

\def\aaref@jnl#1{{\jnl@style#1}}

\def\aj{\aaref@jnl{AJ}}                   
\def\apj{\aaref@jnl{ApJ}}                 
\def\apjl{\aaref@jnl{ApJ}}                
\def\apjs{\aaref@jnl{ApJS}}               
\def\apss{\aaref@jnl{Ap\&SS}}             
\def\aap{\aaref@jnl{A\&A}}                
\def\aapr{\aaref@jnl{A\&A~Rev.}}          
\def\aaps{\aaref@jnl{A\&AS}}              
\def\mnras{\aaref@jnl{Mon.~Not.~Roy.~Astron.~Soc.}}             
\def\prd{\aaref@jnl{Phys.~Rev.~D}}        
\def\prc{\aaref@jnl{Phys.~Rev.~C}}  
\def\prl{\aaref@jnl{Phys.~Rev.~Lett.}}    
\def\qjras{\aaref@jnl{QJRAS}}             
\def\skytel{\aaref@jnl{S\&T}}             
\def\ssr{\aaref@jnl{Space~Sci.~Rev.}}     
\def\zap{\aaref@jnl{ZAp}}                 
\def\nat{\aaref@jnl{Nature}}              
\def\aplett{\aaref@jnl{Astrophys.~Lett.}} 
\def\apspr{\aaref@jnl{Astrophys.~Space~Phys.~Res.}} 
\def\physrep{\aaref@jnl{Phys.~Rep.}}      
\def\physscr{\aaref@jnl{Phys.~Scr}}       
\def\commat{\aaref@jnl{Comm.~Math.~Phys.}}              
\def\science{\aaref@jnl{Science}}               
\def\cqg{\aaref@jnl{Classical Quant.~Grav.}}            
\def\jpcs{\aaref@jnl{JPCS}}                                     
\def\ijmpd{\aaref@jnl{Int.~J.~Mod.~Phys.~D}}                    
\def\grg{\aaref@jnl{Gen.~Relat.~Gravit.}}               
\def\rpp{\aaref@jnl{Rep.~Prog.~Phys.}}          
\def\npa{\aaref@jnl{Nucl.~Phys.~A}}        
\def\lrr{\aaref@jnl{Living Rev.~Rel.}}                   
\def\jcap{\aaref@jnl{J.~Cosmology Astropart.~Phys.}}    
\def\rmp{\aaref@jnl{Rev.~Mod.~Phys.}}   
\def\epjc{\aaref@jnl{Eur.~Phys.~J.~C}}

\allowdisplaybreaks[1]

\begin{document}

\title{A complete study of conformally flat pseudo-symmetric spacetimes in the theory of $F(R)$-gravity}

\author{Avik De\orcidlink{0000-0001-6475-3085}}
\email{de.math@gmail.com}
\affiliation{Department of Mathematical and Actuarial Sciences, Universiti Tunku Abdul Rahman, Jalan Sungai Long, 43000 Cheras, Malaysia}
\author{Simran Arora\orcidlink{0000-0003-0326-8945}}
\email{dawrasimran27@gmail.com}
\affiliation{Department of Mathematics, Birla Institute of Technology and
Science-Pilani,\\ Hyderabad Campus, Hyderabad-500078, India.}
\author{Uday Chand De\orcidlink{0000-0002-8990-4609}}
\email{uc\_de@yahoo.com}
\affiliation{Department of Pure Mathematics, University of Calcutta, West Bengal, India}
\author{P.K. Sahoo\orcidlink{0000-0003-2130-8832}}
\email{pksahoo@hyderabad.bits-pilani.ac.in}
\affiliation{Department of Mathematics, Birla Institute of Technology and
Science-Pilani,\\ Hyderabad Campus, Hyderabad-500078, India.}

\date{\today}
\begin{abstract}
The motive in this article is twofold. First we investigate the geometrical structures of a pseudo-symmetric spacetime $(PS)_4$ with a timelike vector under the condition of conformal flatness. We classify it into two possible types: constant Ricci scalar and closed velocity vector. Then we further study this spacetime as a solution of $F(R)$-gravity theory and show that the pressure and energy density of the effective cosmological perfect fluid separately possess certain typical relations with the geometry and the gravity sector of the theory. Based on this result, some observational and cosmological analyses are done for the different $F(R)$-gravity models and the energy conditions are investigated accordingly, supporting the accelerated expansion of the universe.  
\end{abstract}

\maketitle

\section{Introduction}\label{sec1}
To efficiently reason the late time accelerated expansion of the universe ignoring the assumption of the yet undetected existence of dark energy, researchers tried to modify the Einstein's field equations (EFE). One of the most popular and established such modified theories of gravity is the $F(R)$-theory. By replacing the Ricci scalar $R$ in the Einstein-Hilbert action with an arbitrary function $F(R)$ of $R$ and then varying the action term 
\[S=\frac{1}{\kappa}\int F(R) \sqrt{-g}d^4x +\int L_m\sqrt{-g}d^4x,\]
with respect to $g^{\mu\nu}$ gives the required field equations of $F(R)$-gravity
\begin{equation}
F_R(R)R_{ij}-\frac{1}{2}F(R)g_{ij}+(g_{ij}\Box-\nabla_i\nabla_j)F_R(R)=\kappa T_{ij},\label{FR}
\end{equation}
where $\Box=\nabla^l\nabla_l$. The stress-energy tensor is obtained from the matter Lagrangian $L_m$ by $T_{ij}=-\frac{2}{\sqrt{-g}}\frac{\delta(\sqrt{-g}L_m)}{\delta g^{ij}}$. Viability of $F(R)$-gravity models are extensively studied in the literature \cite{Capozziello/2008,Martino/2014,Sotiriou/2010}. The freedom in developing different functional forms of $F(R)$ gives rise to the challenge of how to confine these numerous possibilities from theoretical or observational features, even though these theories provide an alternate way to explain the cosmic speed up. Additional constraints to $F(R)$ theories may also come by imposing the so-called energy conditions. Recently, some solutions of $F(R)$-gravity has been given with the spacetime geometry restricted to almost pseudo-Ricci symmetric \cite{De/2021}, weakly Ricci symmetric \cite{Avik/2021}, generalized Ricci recurrent \cite{Avik De/2021} type, among others. In fact, imposing these kind of symmetries into the spacetime geometry is always used in general relativity due to the complications involved to solve the system of non-linear partial differential equations. Study of Ricci calculus has a rich history perhaps as old as the theory of general relativity. The study presumably started with locally symmetric manifolds ($\nabla_iR_{hjkl}=0$) and continues to countless newer structures with the development in Physics, specially in the theory of gravity. Chaki \cite{chaki} introduced the notion of pseudo-symmetric manifold $(PS)_n$ as a non-flat Riemannian manifold $M^n, \, n\geq 2$, whose Riemannian curvature tensor satisfies the relation
\begin{equation}
\nabla_lR_{hijk}=2E_lR_{hijk}+E_hR_{lijk}+E_iR_{hljk}+E_jR_{hilk}+E_kR_{hijl}\label{ps}
\end{equation}
where $E_i$ is a non-zero 1-form. A pseudo-Riemannian manifold with Lorentzian type metric tensor of signature $(-,+,+,+)$ is a pseudo-symmetric spacetime if $R_{ijkl}$ satisfies (\ref{ps}). 

Incidentally, in the literature, there are two different notions of pseudo-symmetric manifolds, the one we considered, by Chaki \cite{chaki} and the other by Deszcz \cite{deszcz, ps/2004}, and there is a necessary and sufficient condition for these two concepts to be equivalent \cite{shaikh}. Researchers studied the curvature properties and physical implications of $(PS)_n$ in different settings, including the vanishing Weyl conformal curvature tensor. In \cite{sarbari} a $(PS)_4$ spacetime was studied as the  solution of the standard theory of gravity and it was shown that if the Ricci scalar of such a spacetime is nonzero constant, then the expansion scalar and the acceleration vector of the perfect fluid type matter content vanish. Very recently, it is showed that a pseudo-symmetric generalized Robertson-Walker spacetime is a perfect fluid spacetime and a conformally flat $(PS)_n$ is a generalized Robertson-Walker spacetime \cite{dezhao}. This rejuvenates the research interests in $(PS)_n$ spacetimes with its cosmological implications and this time within the realm of the standard theory of gravity to study its interactions with modified theories of gravity.

The perfect fluid type stress-energy tensor is given by $T_{ij}=(p+\rho)u_iu_j+pg_{ij}$, where $p$ and $\rho$ denote the isotropic pressure and energy density, respectively and $u^i$ is the four-velocity vector of the fluid. In the present discussion we consider a perfect fluid with a four-velocity vector $u^i$ identical to $E^i$. The isotropic pressure $p$ and the energy density $\rho$ of the perfect fluid satisfies a simplified relation $p=w\rho$, where $w$ is called the equation of state (EoS) parameter. The evolution of the energy density and the expansion of the universe is significantly related to the EoS parameter. We can observe different phases of the universe using EoS \cite{Saini/2000}. The dust phase, at $\omega=0$, the radiation-dominated era when $\omega = \frac{1}{3}$, while $\omega=-1$ corresponds to the vacuum energy, i.e. $\Lambda$CDM model. Besides, the accelerating phase of the universe, which has recently been in discussion is portrayed when $\omega < -\frac{1}{3}$. It includes the quintessence phase when $-1 < \omega <0$ and phantom regime $ \omega < -1$. 

In modified gravity, energy conditions define the attractive behavior, spacetime geodesic, and casual structure. Further in physical scenario, energy conditions are also fundamental tools to study black holes and wormholes in different modified gravity \cite{Bamba/2017,Yousaf/2017,Ovgun/2017,Ovgun/2016}  The well-known Raychaudhuri equations are used to explain an attractive aspect of gravity and positive energy density. The null energy condition (NEC), weak energy condition (WEC), dominant energy condition (DEC), and strong energy condition (SEC) are the four primary conditions. These are obtained from Raychaudhuri equations, which allow us to study the complete spacetime structure crucial for understanding cosmological gravitational interactions. The NEC is helpful in discussing the second law of black hole thermodynamics and its violation causes the universe to collapse into a Big-Rip singularity \cite{Carroll/2004}. Under modified theories of gravity, SEC is good at characterizing repulsive or attractive aspects of gravity. The observed accelerated expansion of the universe is supported by the violation of SEC. Some of the works on energy conditions in $F(R)$ gravity are studied in \cite{Atazadeh/2009,Santos/2010}. Also, the study of energy conditions using the mathematical framework known as cosmography enables us to bound the free parameters constraining the families of different gravity models compatible with the accelerated expansion of the universe \cite{Mandal/2020, Arora/2021}. The energy conditions enable us to specify our free parameters, limiting the families of of $F(R)$ models that are compatible with the  accelerated expansion of the universe. 

The present paper is organized as follows:
We consider a 4-dimensional pseudo-symmetric spacetime and examine the equation of motion in section \ref{sec2} after introducing the necessary ingredients of the present study in the first section. In section \ref{sec3}, we observe the behavior of cosmological parameters such as density parameter, the equation of state parameter. Further, various energy conditions are investigated in section \ref{sec4}. The discussions and conclusion are presented in section \ref{sec5}.

Throughout the article we use the notations $\dot{R}=E^k\nabla_kR,\, \ddot{R}=E^k\nabla_k\dot{R},\, \dot{E_i}=E^k\nabla_kE_i$.

\section{The equations of motion}\label{sec2}

In this section we consider a 4-dimensional pseudo-symmetric spacetime, hence its $(0,4)$- type Riemannian curvature tensor $R_{ijkl}$ satisfies (\ref{ps}). Contracting simultaneously $h$, $k$ and $j$, $l$ in (\ref{ps}) we respectively obtain 
\begin{equation}
\nabla_lR_{ij}=2E_lR_{ij}+E_iR_{lj}+E_jR_{il}+E^kR_{lijk}+E^kR_{ljik}\label{u1}
\end{equation}
and
\begin{equation}
\nabla_iR=4E^lR_{il}+2RE_i\label{nabr}.
\end{equation}
Transvecting (\ref{nabr}) by $E^i$ we get
\begin{equation}
\dot{R}+2R=4E^iE^lR_{il}.\label{u2}
\end{equation}
Now, from the contracted Bianchi's second identity we have the well-known relation \cite{hawking}
\begin{equation}
\nabla^mR_{jklm}=\nabla_jR_{kl}-\nabla_kR_{jl}.\label{a1}
\end{equation}
Contracting $k$ and $l$ in (\ref{ps}) and using (\ref{a1}) we obtain
\begin{equation}
\nabla_hR_{ij}-\nabla_iR_{hj}=3E^kR_{hijk}+E_hR_{ij}-E_iR_{hj}.\label{u3}
\end{equation}
Since we consider a flat conformal curvature tensor, we also have vanishing conformal divergence in this spacetime, which gives the relation \cite{hawking}
\begin{equation}
6[\nabla_iR_{jk}-\nabla_jR_{ik}]=g_{jk}\nabla_iR-g_{ik}\nabla_jR.\label{conf}
\end{equation}
Using (\ref{nabr}), (\ref{u3}) and (\ref{conf}) we get
\begin{equation}
9E^kR_{hijk}+3E_hR_{ij}-3E_iR_{hj}\\=g_{ij}[RE_h+2E^lR_{hl}]-g_{hj}[RE_i+2E^lR_{il}].\label{u4}
\end{equation}
Now a conformally flat spacetime secures a special format of the Riemannian curvature tensor expressed in terms of the Ricci curvature tensor and the Ricci scalar \cite{hawking}
\begin{equation}
6R_{hijk}=(3R_{ij}-Rg_{ij})g_{hk}-(3R_{hj}-Rg_{hj})g_{ik}\\+3(g_{ij}R_{hk}-g_{hj}R_{ik}).\label{u6}
\end{equation}
Transvecting (\ref{u6}) by $E^k$ and using (\ref{u4}) we obtain
\begin{equation}
E_h(9R_{ij}-3Rg_{ij})-E_i(9R_{hj}-3Rg_{hj})+9E^k(g_{ij}R_{hk}-g_{hj}R_{ik})\\ +6E_hR_{ij}-6E_iR_{hj}=g_{ij}[2RE_h+4E^lR_{lh}]-\\
g_{hj}[2RE_i+4E^lR_{il}]
\end{equation}
which simplifies to
\begin{equation}
3E_hR_{ij}-RE_hg_{ij}-3E_iR_{hj}+RE_ig_{hj}+g_{ij}E^lR_{lh}-g_{hj}E^lR_{il}=0.\label{u7}
\end{equation}
Transvecting (\ref{u7}) by $E^h$ we obtain
\begin{equation}
3R_{ij}=Rg_{ij}+RE_iE_j+g_{ij}E^kE^lR_{kl}-3E_iE^lR_{jl}-E_jE^lR_{il}.\label{rij}
\end{equation}
On the other hand, transvecting (\ref{u4}) by $E^j$ we obtain 
\begin{equation}
E_hE^lR_{il}=E_iE^lR_{hl}\label{v5}.
\end{equation}
And hence by (\ref{u2}) we have
\begin{equation}
4E^lR_{il}=-4E_iE^hE^lR_{hl}=-E_i[\dot{R}+2R]\label{u8}.
\end{equation}
(\ref{v5}) and (\ref{u8}) reduces (\ref{rij}) to
\begin{equation}
R_{ij}=\frac{[6R+\dot{R}]}{12}g_{ij}+\frac{[3R+\dot{R}]}{3}E_iE_j.\label{rij2}
\end{equation}
\subsubsection{\bf{$\dot{R}+3R=0$}}

In this particular case, the discussed conformally flat spacetime reduces to an Einstein manifold by the virtue of (\ref{rij2}) and thus turns out to be a spaceform itself. Also, on integration of $\dot{R}+3R=0$ we obtain $R=R_0e^{-3t}$ where by $R_0$ we denote the initial value of the Ricci scalar. 

\subsubsection{\bf{$\dot{R}+3R\neq0$}}
Covariantly differentiating (\ref{rij2}) and then transvecting by $E^j$ we get
\begin{equation}\label{u9}
4[\dot{R}+3R]\nabla_kE_i=-3[\nabla_k\dot{R}+2\nabla_kR]E_i-12E^j\nabla_kR_{ij}
\end{equation}
Transvecting (\ref{u9}) by $E^k$ we obtain
\begin{equation}  4(\dot{R}+3R)\dot{E_i}=-3(\ddot{R}+2\dot{R})E_i-12E^jE^k\nabla_kR_{ij}.\label{k10} \end{equation}
From (\ref{u1}) and (\ref{u8}) we obtain
\begin{equation} E^jE^k\nabla_kR_{ij}=(\dot{R}+2R)E_i.\label{Aa}\end{equation}
From (\ref{k10}) and (\ref{Aa}) we get
\begin{equation} 4(\dot{R}+3R)\dot{E_i}=(-3\ddot{R}-18\dot{R}-24R)E_i\end{equation}
which by $E_i\dot{E^i}=0$, readily gives us the acceleration vector 
\begin{equation} \dot{E_i}=0.\label{aA}\end{equation}
Whereas using (\ref{u8}) from (\ref{nabr}) we obtain the correspondence
\begin{equation}
\nabla_iR-2RE_i=-\dot{R}E_i-2RE_i
\end{equation}
which provides
\begin{equation}
\nabla_iR=-\dot{R}E_i.\label{nabr2}
\end{equation}
Differentiate (\ref{nabr2}) covariantly we obtain
\begin{equation}
\nabla_j\nabla_iR=-\dot{R}\nabla_jE_i-E_i\nabla_j\dot{R}.\label{nabijr1}
\end{equation}
Swapping $i$  and $j$ in (\ref{nabijr1}) we obtain
\begin{equation}
\nabla_i\nabla_jR=-\dot{R}\nabla_iE_j-E_j\nabla_i\dot{R}.\label{nabijr}
\end{equation}
$R$ being a scalar field, this combination produces
\begin{equation}
\dot{R}(\nabla_jE_i-\nabla_iE_j)=E_j\nabla_i\dot{R}-E_i\nabla_j\dot{R}.
\end{equation}
Transvecting the above equation by $E^j$ we get
\begin{equation}
\nabla_i\dot{R}=-E_i\ddot{R}
\end{equation}
which transforms (\ref{nabijr1}) and (\ref{nabijr}) respectively into
\begin{equation}
\nabla_j\nabla_iR=-\dot{R}\nabla_jE_i+\ddot{R}E_iE_j
\end{equation}
\begin{equation}
\nabla_i\nabla_jR=-\dot{R}\nabla_iE_j+\ddot{R}E_iE_j.\label{nabijr2}
\end{equation}
But these two equations together imply
\begin{equation}
\dot{R}[(\nabla_jE_i-\nabla_iE_j))]=0
\end{equation}
So, either $\dot{R}=0$ which by (\ref{nabr2}) implies a constantor $R$, or $(\nabla_jE_i-\nabla_iE_j)=0$, that is, $E$ is a closed 1-form. We investigate these two cases in two separate subsections.

\subsection{Case I: Closed 1-form $E$}

In this subsection we consider a non-constant $R$. (\ref{u1}) by (\ref{u2}), (\ref{u8}) and  (\ref{rij2}) gives
\begin{equation} 
E^j\nabla_kR_{ij}=-\left[\frac{13}{12}\dot{R}+\frac{5}{2}R\right]E_iE_k-\\
\frac{1}{12}(6R+\dot{R})g_{ik}+E^lE^jR_{kjil}\label{u10}
\end{equation}
and, (\ref{u6}) by (\ref{u2}), (\ref{u8}) and (\ref{rij2}) gives
\begin{equation}
6E^lE^jR_{kjil}=-(R+\frac{\dot{R}}{2})g_{ik}-(R+\frac{\dot{R}}{2})E_iE_k.\label{u11}\end{equation} 
Combining (\ref{u9}),(\ref{u10}),(\ref{u11}), we obtain
\begin{equation} 
\nabla_kE_i=\frac{\dot{R}+4R}{2(\dot{R}+3R)}g_{ik}+\frac{3\ddot{R}+20\dot{R}+32R}{4(\dot{R}+3R)}E_iE_k.\label{u13}
\end{equation}
Therefore, using (\ref{u13}), from (\ref{nabijr2}) we finally conclude that
\begin{equation}
\nabla_i\nabla_jR=\frac{\dot{R}\ddot{R}+12R\ddot{R}-32R\dot{R}-20\dot{R}^2}{4(\dot{R}+3R)}E_iE_j-\frac{\dot{R}^2+4R\dot{R}}{2(\dot{R}+3R)}g_{ij}
\end{equation}
and 
\begin{equation}
\Box R=\frac{12\dot{R}^2-12R\ddot{R}-\dot{R}\ddot{R}}{4(\dot{R}+3R)}.
\end{equation}
We summarise these results as:\\

\textbf{Theorem 2.1}

In a conformally flat $(PS)_4$ spacetime with non-constant Ricci scalar, if $\dot{R}+3R\neq0$, we have the followings: \\
\begin{enumerate}
\item $R_{ij}=\frac{[6R+\dot{R}]}{12}g_{ij}+\frac{[3R+\dot{R}]}{3}E_iE_j.$
\item $E^l\nabla_lE_i=0.$
\item $\nabla_iR=-\dot{R}E_i.$
\item $\nabla_kE_i=\frac{\dot{R}+4R}{2(\dot{R}+3R)}g_{ik}+\frac{3\ddot{R}+20\dot{R}+32R}{4(\dot{R}+3R)}E_iE_k.$
\item $\nabla_i\nabla_jR=\frac{\dot{R}\ddot{R}+12R\ddot{R}-32R\dot{R}-20\dot{R}^2}{4(\dot{R}+3R)}E_iE_j-\frac{\dot{R}^2+4R\dot{R}}{2(\dot{R}+3R)}g_{ij}.$
\item $\Box R=\frac{12\dot{R}^2-12R\ddot{R}-\dot{R}\ddot{R}}{4(\dot{R}+3R)}.$
\end{enumerate}
Now, for any analytic function $F(R)$ we have
\begin{equation}
\nabla_i\nabla_jF_R(R)=F_{RR}(R)\nabla_i\nabla_jR+F_{RRR}(R)(\nabla_iR)(\nabla_jR).
\end{equation}
Using the curvature properties in Theorem 2.1 we conclude that\\

\textbf{Theorem 2.2}

In a conformally flat $(PS)_4$ spacetime with non-constant Ricci scalar, if $\dot{R}+3R\neq0$, for any analytic function $F(R)$ we have
\begin{widetext}
\begin{equation}
\nabla_i\nabla_jF_R(R)=\left[\left(\frac{\dot{R}\ddot{R}+12R\ddot{R}-32R\dot{R}-20\dot{R}^2}{4(\dot{R}+3R)}\right)F_{RR}(R)+\dot{R}^2F_{RRR}(R)\right]E_iE_j 
-\frac{\dot{R}^2+4R\dot{R}}{2(\dot{R}+3R)}F_{RR}(R)g_{ij} 
\end{equation}
\end{widetext}
and
\begin{equation}
\Box F_R(R)=\frac{12\dot{R}^2-12R\ddot{R}-\dot{R}\ddot{R}}{4(\dot{R}+3R)}F_{RR}(R)-\dot{R}^2F_{RRR}(R)
\end{equation}

If a conformally flat $(PS)_4$ is a solution of the $F(R)$-gravity equations (\ref{FR}), using Theorem 2.2 we can express 
\begin{widetext}
\begin{multline}
 \kappa T_{ij}=\left[ -\frac{F(R)}{2}+\frac{\dot{R}+6R}{12}F_R(R)+\frac{14\dot{R}^2+8R\dot{R}-12R\ddot{R}-\dot{R}\ddot{R}}{4(\dot{R}+3R)}F_{RR}(R)-\dot{R}^2F_{RRR}(R)\right]g_{ij}
\\+\left[\frac{\dot{R}+3R}{3}F_R(R)-\frac{\dot{R}\ddot{R}+12R\ddot{R}-32R\dot{R}-20\dot{R}^2}{4(\dot{R}+3R)}F_{RR}(R)-\dot{R}^2F_{RRR}(R)\right]E_iE_j,\label{a}
\end{multline}
\end{widetext}
which implies that $T_{ij}$ denotes a perfect fluid energy-momentum tensor with isotropic pressure given by 
\begin{widetext}
\begin{equation}
\kappa p=-\frac{F(R)}{2}+\frac{\dot{R}+6R}{12}F_R(R)+\frac{14\dot{R}^2+8R\dot{R}-12R\ddot{R}-\dot{R}\ddot{R}}{4(\dot{R}+3R)}F_{RR}(R)-\dot{R}^2F_{RRR}(R)
\end{equation}
 \end{widetext}
and the energy density given by

\begin{equation}
\kappa\rho=\frac{F(R)}{2}+\frac{\dot{R}+2R}{4}F_R(R)-\frac{3\dot{R}(\dot{R}+4R)}{2(\dot{R}+3R)}F_{RR}(R).
 \end{equation}

This leads to one of the main results of the present study:\\

\textbf{Theorem 2.3}

In a conformally flat $(PS)_4$ ($\dot{R}+3R\neq0$) satisfying the theory of $F(R)$-gravity with a non-rotating matter content is a perfect fluid with four-velocity vector $E^i$; energy density 
\begin{equation}
\rho=\frac{\frac{F(R)}{2}+\frac{\dot{R}+2R}{4}F_R(R)-\frac{3\dot{R}(\dot{R}+4R)}{2(\dot{R}+3R)}F_{RR}(R)}{\kappa}
\end{equation}
and isotropic pressure 
\begin{equation}
p=\frac{-\frac{F(R)}{2}+\frac{\dot{R}+6R}{12}F_R(R)+\frac{14\dot{R}^2+8R\dot{R}-12R\ddot{R}-\dot{R}\ddot{R}}{4(\dot{R}+3R)}F_{RR}(R)-\dot{R}^2F_{RRR}(R)}{\kappa}
\end{equation}

\textbf{Remark 2.1}

In general relativity, $F(R)=R$, so the perfect fluid in that case has isotropic pressure $p=\frac{\dot{R}}{12\kappa}$ and the energy density $\rho=\frac{R+\dot{R}/4}{\kappa}$.

It is known that a Lorentzian manifold $M$ of dimension $n>3$ with Ricci curvature tensor $R_{kl}=Ag_{kl}+Bu_ku_l$, where $A$ and $B$ are scalar fields, $u$ is a closed timelike unit vector field and with a conformally harmonic curvature tensor is a generalized Robertson-Walker (GRW) spacetime with a Einstein fibre \cite{ucdpf}. Moreover, a GRW spacetime $M$ is conformally flat if and only if its fibre is a space of constant curvature, that is $M$ is Robertson-Walker (RW) \cite{rwsf}. Using (\ref{rij2}) therefore we assert that\\

\textbf{Theorem 2.4}

A conformally flat $(PS)_4$ with a non-constant Ricci scalar $R$ is a Robertson-Walker spacetime.

\subsection{Case II: constant $R$} 

In this case, from (\ref{nabr}) we get 
\begin{equation}
2E^lR_{il}=-RE_i.\label{b1}
\end{equation}
For a constant $R$, the field equations of $F(R)$-gravity (\ref{FR}) reduces to
\begin{equation}
R_{ij}-\frac{R}{2}g_{ij}=\frac{\kappa}{F_R(R)}T^{\text{eff}}_{ij},\label{frcon}
\end{equation}
where $T^{\text{eff}}_{ij}=T_{ij}+\frac{1}{2k}[F(R)-RF_R(R)]g_{ij}$.
Transvecting (\ref{frcon}) by $E^j$ and using (\ref{b1}) we get
\begin{equation}
R=\frac{2\kappa\rho-F(R)}{F_R(R)}
\end{equation}
which gives 
\begin{equation}
\rho=\frac{RF_R(R)+F(R)}{2\kappa}. \label{rho2}
\end{equation}
Again, the trace of (\ref{frcon}) by the expressions of $R$ and $\rho$ found above gives us
\begin{equation}
p=\frac{RF_R(R)-F(R)}{2\kappa}. \label{p2}
\end{equation}
We state that\\

\textbf{Theorem 2.4}

In a conformally flat $(PS)_4$ (with $\dot{R}=0$) satisfying the theory of $F(R)$-gravity, the matter content is a perfect fluid with four-velocity vector $E^i$; isotropic pressure $p=\frac{RF_R(R)-F(R)}{2\kappa}$ and energy density $\rho=\frac{RF_R(R)+F(R)}{2\kappa}$.\\
In such case, the EoS $w$ is given by
\begin{equation}
 w=\frac{p}{\rho}=1-\frac{2F(R)}{F(R)+RF_R(R)}.
\end{equation}

\textbf{Remark 2.2}

In a conformally flat $(PS)_4$ with constant $R$ satisfying the field equations of $F(R)$-gravity with $F_R(R)>0$, we readily have the followings:\\
(i) For a positive Ricci scalar $R>0$, $w>-1$.\\
(ii)For a negative Ricci scalar $R<0$, $w<-1$, a hypothetical phantom energy state.


\section{Cosmological Study} \label{sec3}

The Hubble parameter $H=\frac{\dot{a}}{a}$, where $a$ is the scale factor and $(\cdot)$ represent the derivative with respect to time, is a cosmological parameter extensively used to study the rate of expansion of the universe. Besides, the expansion of scale factor with respect to cosmic time yields the higher order derivatives. The first four derivatives are known as velocity, acceleration $(q)$, jerk $(j)$ and snap $(s)$ parameters respectively \cite{Visser/2004}. This cosmography is a mathematical tool for the description of the universe. The current values of these parameters with subscript $0$ represent the evolution of the universe. That is, $q_{0}<0$ signifies an accelerated expansion, whereas $j_{0}$ and $s_{0}$ distinguish between different accelerating models.

We first note that the Ricci scalar and its derivatives for a spatially flat FLRW geometry can be expressed in terms of the deceleration ($q$), jerk ($j$), and snap ($s$) parameters \cite{Santos/2007},
\begin{align} \label{R}
R &= -6 H^{2} (1-q)\\ 
\label{R1}
\dot{R}&= -6 H^{3} (j-q-2)\\
\label{R2}
\ddot{R}&= -6 H^{4} (s+q^{2}+8q+6)
\end{align}

where 
\begin{align}\label{q}
q &= -\frac{1}{H^{2}} \frac{\ddot{a}}{a} \\ 
\label{j}
j &= \frac{1}{H^{3}} \frac{\dddot{a}}{a} \\ 
\label{s}
s &= \frac{1}{H^{4}} \frac{\ddddot{a}}{a}    
\end{align}

\textbf{Case 2.1.1 : $\dot{R}+3R=0$ }\\
As studied in section 2, we have a solution in this case as $R= R_{0} e^{-3t}$. This form itself describes the space. Since, there is no functional form $f$ in the solution, therefore it can be treated as the General Relativity (GR) case.\\

\textbf{Case 2.1.2 : $\dot{R}+3R \neq 0$ }

We obtain density and pressure in terms of deceleration parameter, jerk and snap parameters using eqs.\eqref{R},\eqref{R1},\eqref{R2} as 

\begin{multline}
 \rho = \frac{1}{\kappa}\left( \frac{F}{2}-\frac{3}{2} F_{R} \left( H^3 (j-q-2)+2 H^2 (1-q)\right) +\frac{3 F_{RR} H^3 (j-q-2) \left(-6 H^3 (j-q-2)-24 H^2 (1-q)\right)}{-2 H^3 (j-q-2)-6 H^2 (1-q)}\right) 
\end{multline} 

\begin{multline}
 p= \frac{1}{\kappa}\left( -\frac{F}{2}-\frac{1}{2} F_{R} \left(H^3 (j-q-2)+6 H^2 (1-q)\right)-36 F_{RRR} H^6 (j-q-2)^2 + \right.\\
\left. \frac{F_{RR} (-36 H^7 (j-q-2) \left(q^2+8 q+s+6\right)+504 H^6 (j-q-2)^2-432 H^6 (1-q) \left(q^2+8 q+s+6\right)+288 H^5 (1-q) (j-q-2))}{4 \left(-6 H^3 (j-q-2)-18 H^2 (1-q)\right)}\right) 
\end{multline}

\subsection{Model-I}

We assume the functional form of $F(R)= R+ \alpha log(\beta R)$ \cite{Girones/2010}, where $\alpha$ and $\beta$ are constants. It is seen that Logarithm function is continuous and differential when $\beta R > 0$. Therefore, we choose $\beta \neq 0$. For $\alpha=0$, the model reduces to the well motivated GR case. We consider the present values of $q_{0}= -0.55$, $j_{0}=1$, $s_{0}= -0.35$ and $H_{0}=67.9 $ km/s/Mpc \cite{Capo/2019, Planck/2018}. Then we can have expressions for density $\rho$ and pressure $p$ as,

\begin{multline}
\displaystyle \rho= \frac{1}{4k} \left( \frac{(H (-j+q+2)+2 (q-1))(\alpha +6 H^2 (q-1))}{q-1}+2 \left(\alpha  \log \left(6 \beta  H^2 (q-1)\right)+6 H^2 (q-1)\right) \right.\\
\left. +\frac{\alpha  (-j+q+2) (H (-j+q+2)+4 (q-1))}{H (q-1)^2 (H (-j+q+2)+3 (q-1))}\right)  
\end{multline}

and 

\begin{multline}
\displaystyle p= -\frac{1}{24 \kappa} \left(  \frac{2 (H (j-q-2)-6 q+6) \left(\alpha +6 H^2 (q-1)\right)}{q-1}+12 \left(\alpha  \log \left(6 \beta  H^2 (q-1)\right)+6 H^2 (q-1)\right) \right. \\
 \left. + \frac{ \alpha \left(H^2 (-(j-q-2)) \left(q^2+8 q+s+6\right)+2 H (7 j^2-14 j (q+2)+6 q^3+49 q^2+6 q s+ 16 q-6 s-8)+8 (-j q+j+q^2+q-2)\right)}{H (q-1)^2 (H (-j+q+2)+3 (q-1))} \right. \\
 \left. +\frac{8 \alpha  (-j+q+2)^2}{(q-1)^3} \right) 
\end{multline}

\begin{figure}[H]
     \centering
     \begin{subfigure}[b]{0.45\textwidth}
         \centering
         \includegraphics[width=\textwidth]{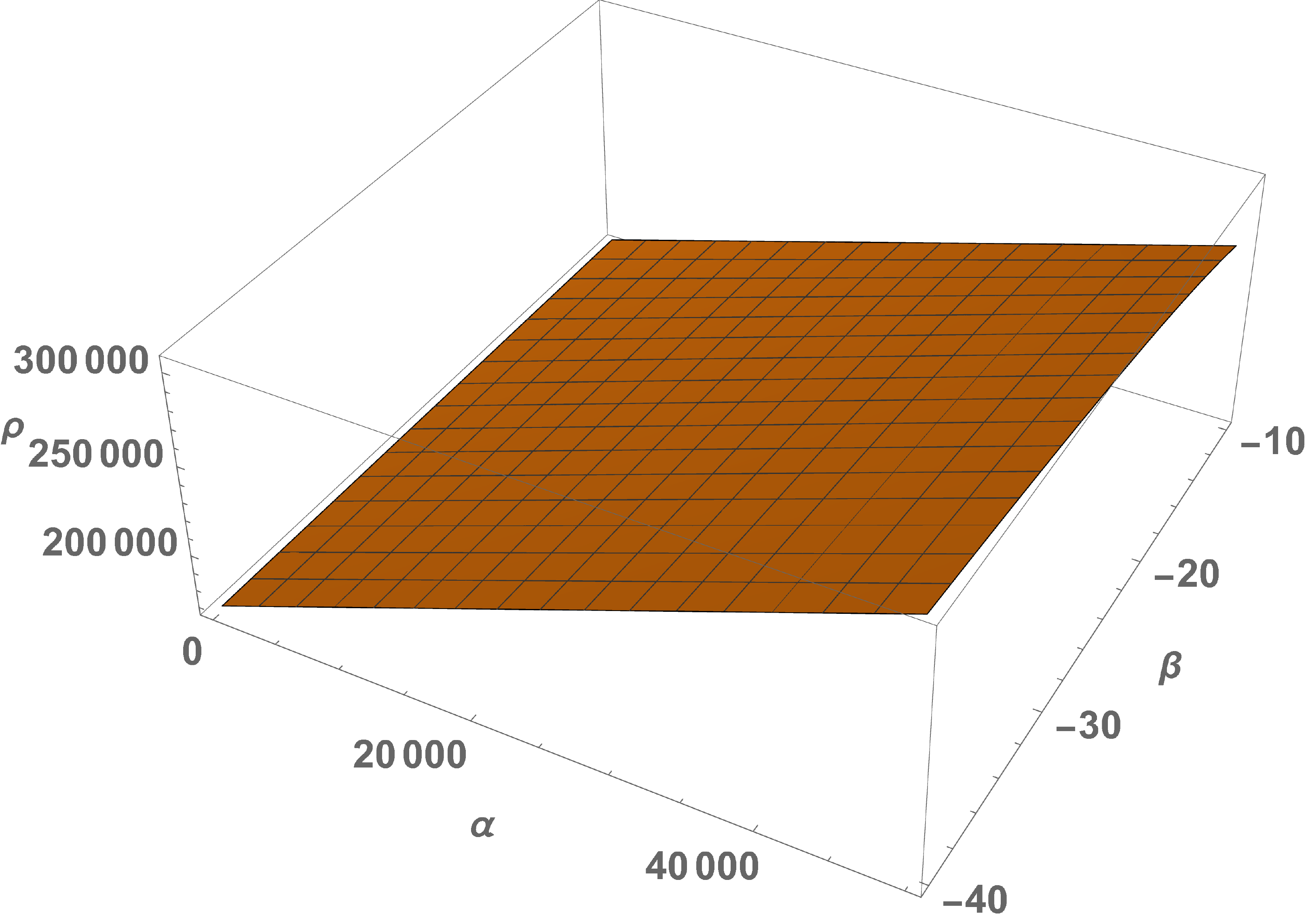}
         \caption{The density parameter.}
         \label{fig:density}
     \end{subfigure}
     \hfill
     \begin{subfigure}[b]{0.45\textwidth}
         \centering
         \includegraphics[width=\textwidth]{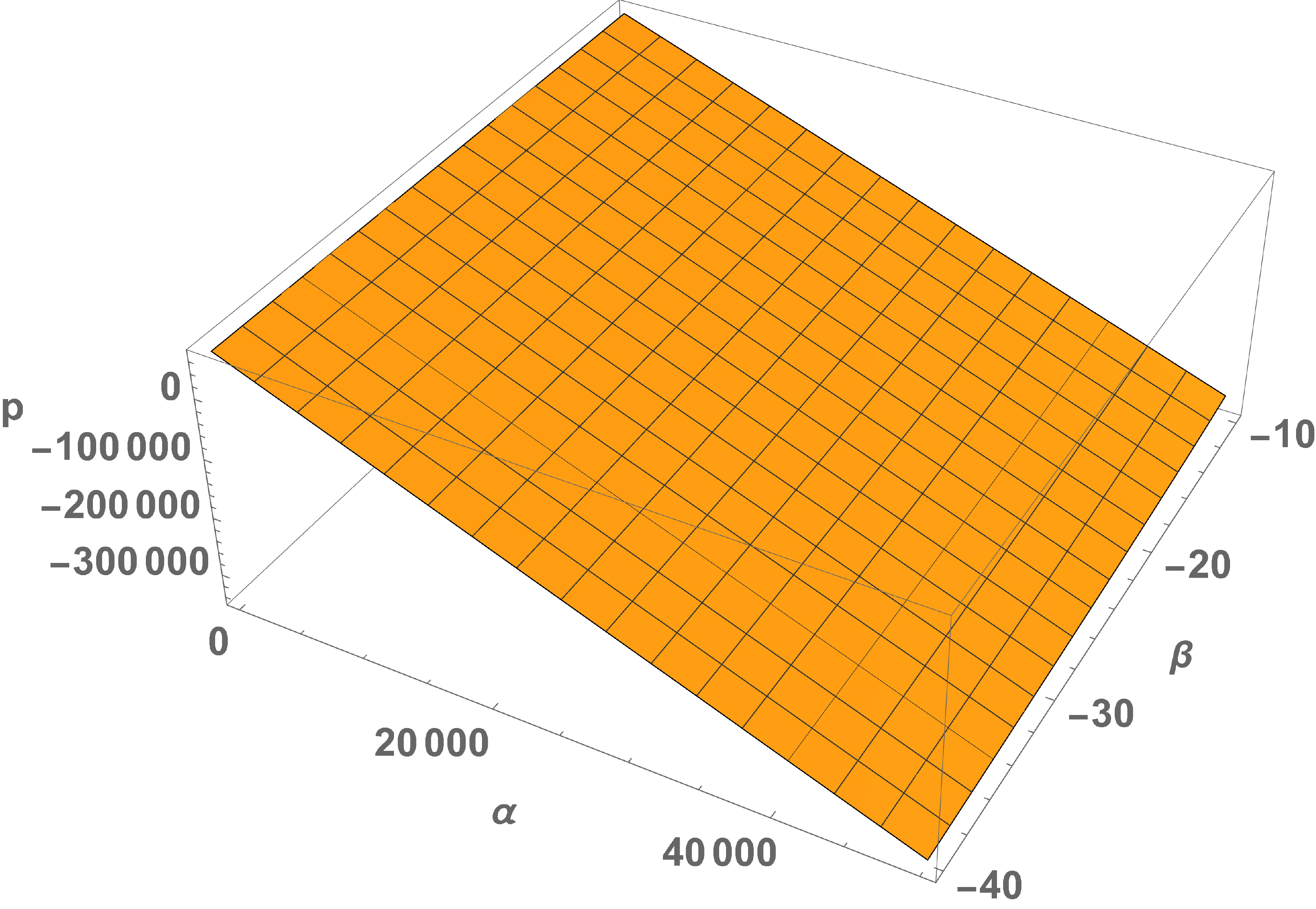}
         \caption{The pressure parameter.}
         \label{fig:pressure}
     \end{subfigure}
     \hfill
     \begin{subfigure}[b]{0.45\textwidth}
         \centering
         \includegraphics[width=\textwidth]{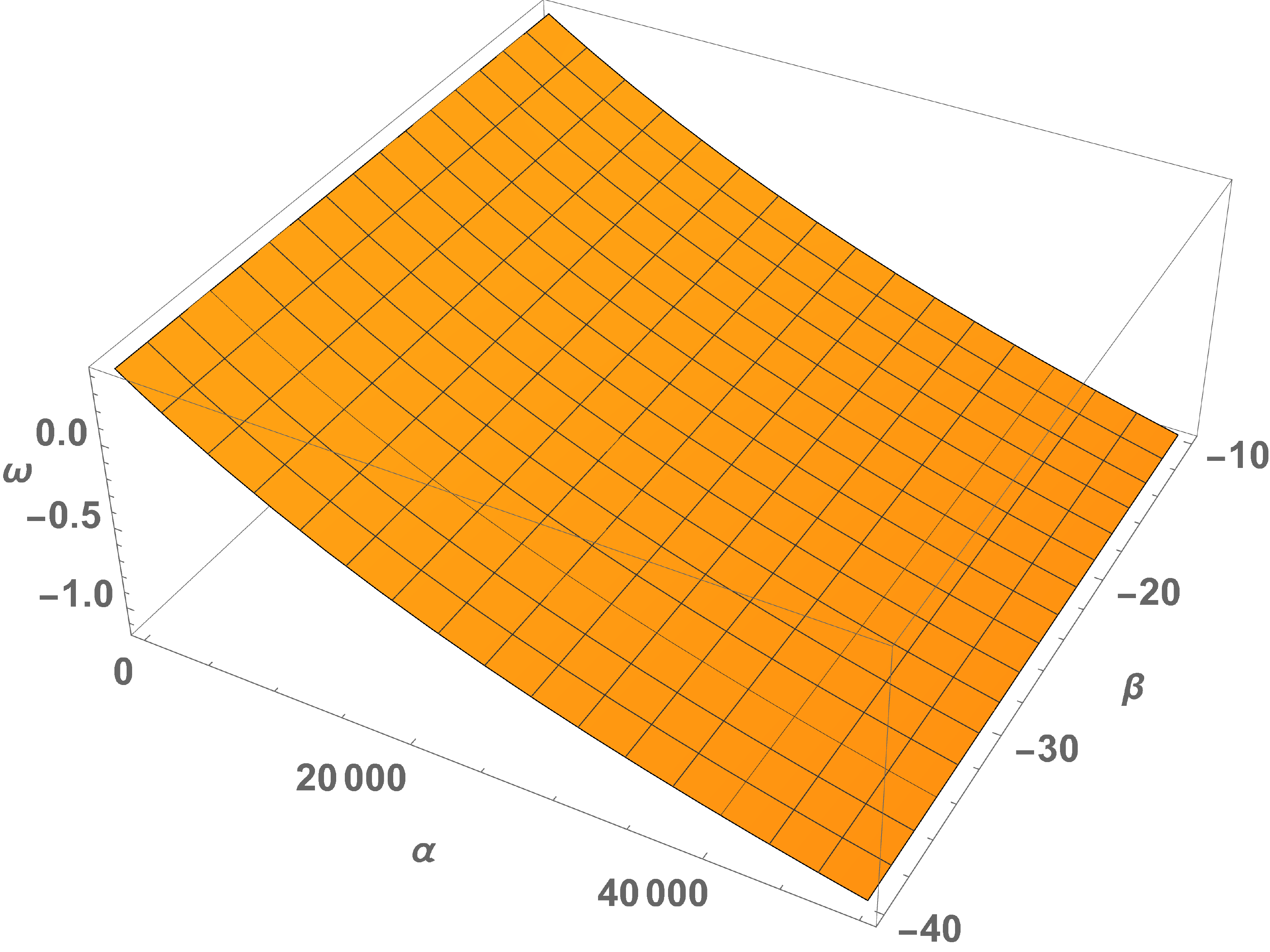}
         \caption{The EoS parameter.}
         \label{fig:omega}
     \end{subfigure}
        \caption{The behavior of the density, pressure and the EoS parameters considering $F(R)= R+ \alpha log(\beta R)$ for Case 2.1.2 with $ -1 \leq \alpha \leq 50000$ and $-40\leq \beta \leq -10$.}
        \label{fig:model1 Case1}
\end{figure}

Since $F(R) = R+ \alpha log(\beta R)$ is defined for $\beta R > 0$ and $R$ is negative with the value of $q_{0}$, therefore we consider $\beta < 0$. Also, $\alpha =0$ corresponds to $F(R)=R$, which reduces to the case of  GR. We obtain the bounds of $\alpha$ through the behavior of energy density $\rho$, which is always positive, as shown in Fig. \ref{fig:density} and negative pressure depicted in Fig. \ref{fig:pressure}. It is readily seen that $\beta$ contributes more to the energy density and $\rho$ increases for more negative $\beta$ and decreases for less negative $\beta$.
Further, the equation of state parameter(EoS) is the relationship between $p$ and $\rho$ i.e., $\omega = \frac{p}{\rho}$. It is used to classify various epochs, such as decelerated to accelerated phases of the universe. So, $\omega =0$ represent the matter-dominated phase, $\omega= \frac{1}{3}$ shows the radiation whereas the accelerated phases are depicted as the quintessence when $-1<\omega <0$, the $\Lambda$CDM as $\omega=-1$ and $\omega <-1$ is the phantom era. As a result, the EoS parameter is thought to be a good choice for comparing our models to $\Lambda$CDM. So, the EoS parameter is used to constrain model parameters to investigate various energy conditions.
The behavior of the EoS parameter for our considered model is shown in Fig. \ref{fig:omega}. The plot of EoS shows a transition from positive to negative supporting an acceleration in the universe and lies in the quintessence phase $-1 < \omega < 0$.

\textbf{Case II : $\dot{R} = 0$ } 

In this case $R$ is constant. We obtain $\rho$ and $p$ for this case in section \ref{sec2}. Using equations \eqref{rho2}, \eqref{p2}, \eqref{R} and the functional form $F(R)=R+\alpha log(\beta R)$, where $\alpha$, $\beta$ are constants. Again, for $\alpha=0$, the model reduces to GR. Further from \eqref{p},   we get $p=0$ for $\alpha=0$, and according  to it $\rho$ is constant which is instant the case of cosmological constant. We assumed $\alpha$ different from zero to analyze the derived model to study different cases rather than the cosmological constant.
\begin{equation}
\rho = \frac{\alpha +\alpha  \log \left(6 \beta  H^2 (q-1)\right)+12 H^2 (q-1)}{2 \kappa}.
\end{equation} 

\begin{equation}
p=\frac{\alpha -\alpha  \log \left(6 \beta  H^2 (q-1)\right)}{2 \kappa}.
\label{p}
\end{equation}

The Eos parameter $\omega=\frac{p}{\rho}$ is obtained as 

\begin{equation}
\omega= 1-\frac{2 \left(\alpha  \log \left(-6 \beta  H^2 (1-q)\right)-6 H^2 (1-q)\right)}{\alpha  \log \left(-6 \beta  H^2 (1-q)\right)-6 H^2 (1-q) \left(1-\frac{\alpha }{6 H^2 (1-q)}\right)-6 H^2 (1-q)}
\end{equation}

\begin{figure}[H]
     \centering
     \begin{subfigure}[b]{0.45\textwidth}
         \centering
         \includegraphics[width=\textwidth]{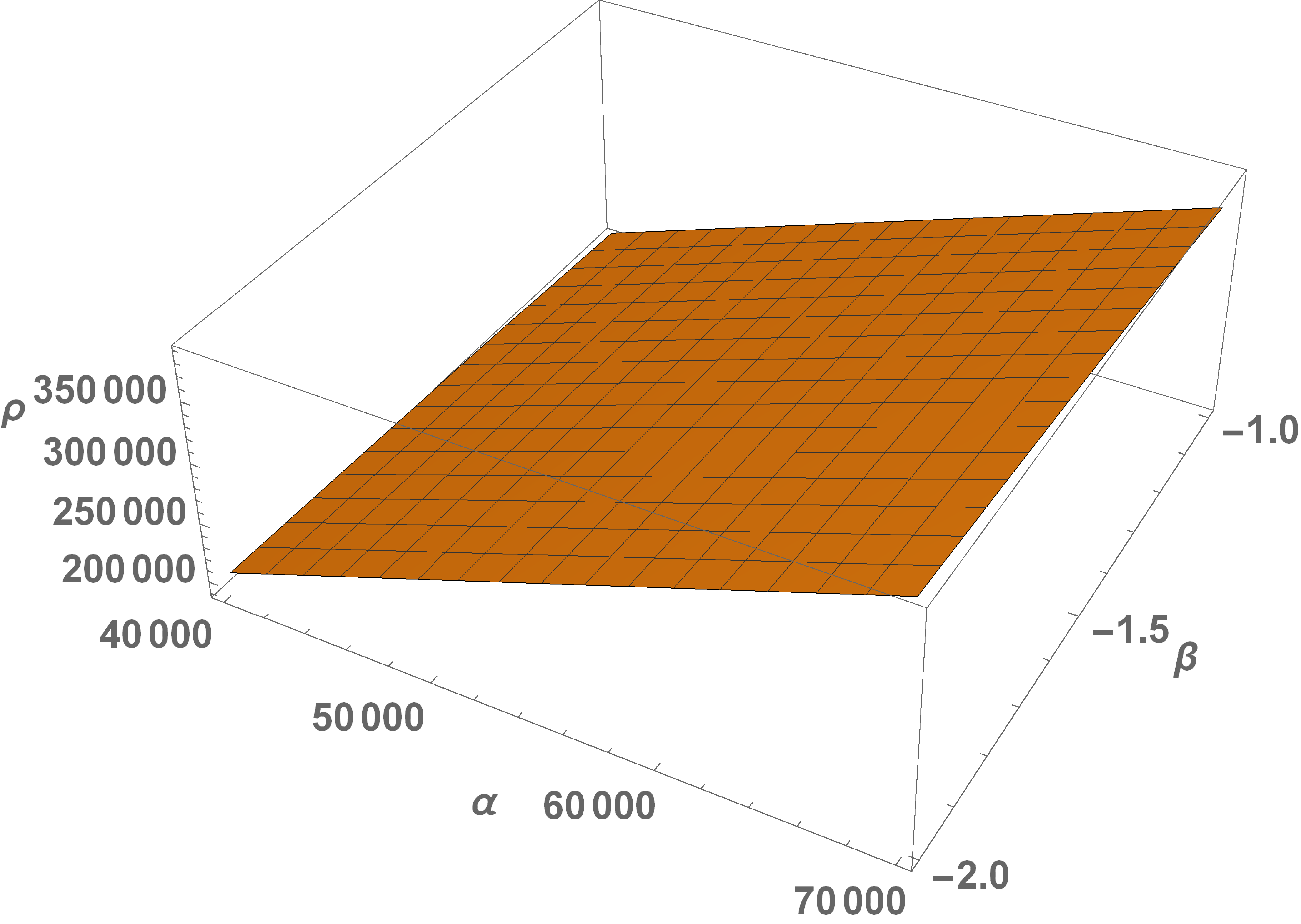}
         \caption{The density parameter.}
         \label{fig:density2}
     \end{subfigure}
     \hfill
     \begin{subfigure}[b]{0.45\textwidth}
         \centering
         \includegraphics[width=\textwidth]{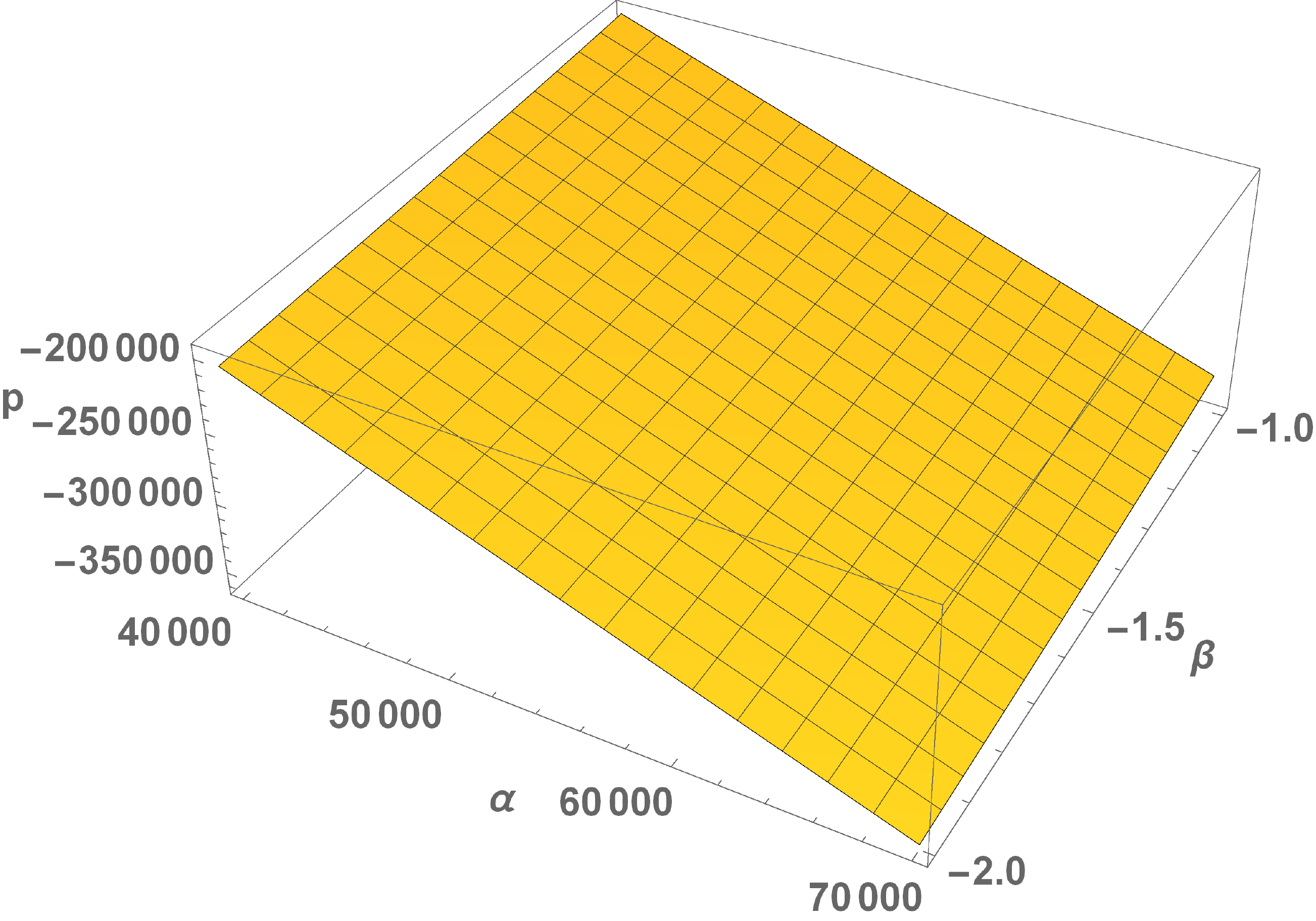}
         \caption{The pressure parameter.}
         \label{fig:pressure2}
     \end{subfigure}
     \hfill
     \begin{subfigure}[b]{0.45\textwidth}
         \centering
         \includegraphics[width=\textwidth]{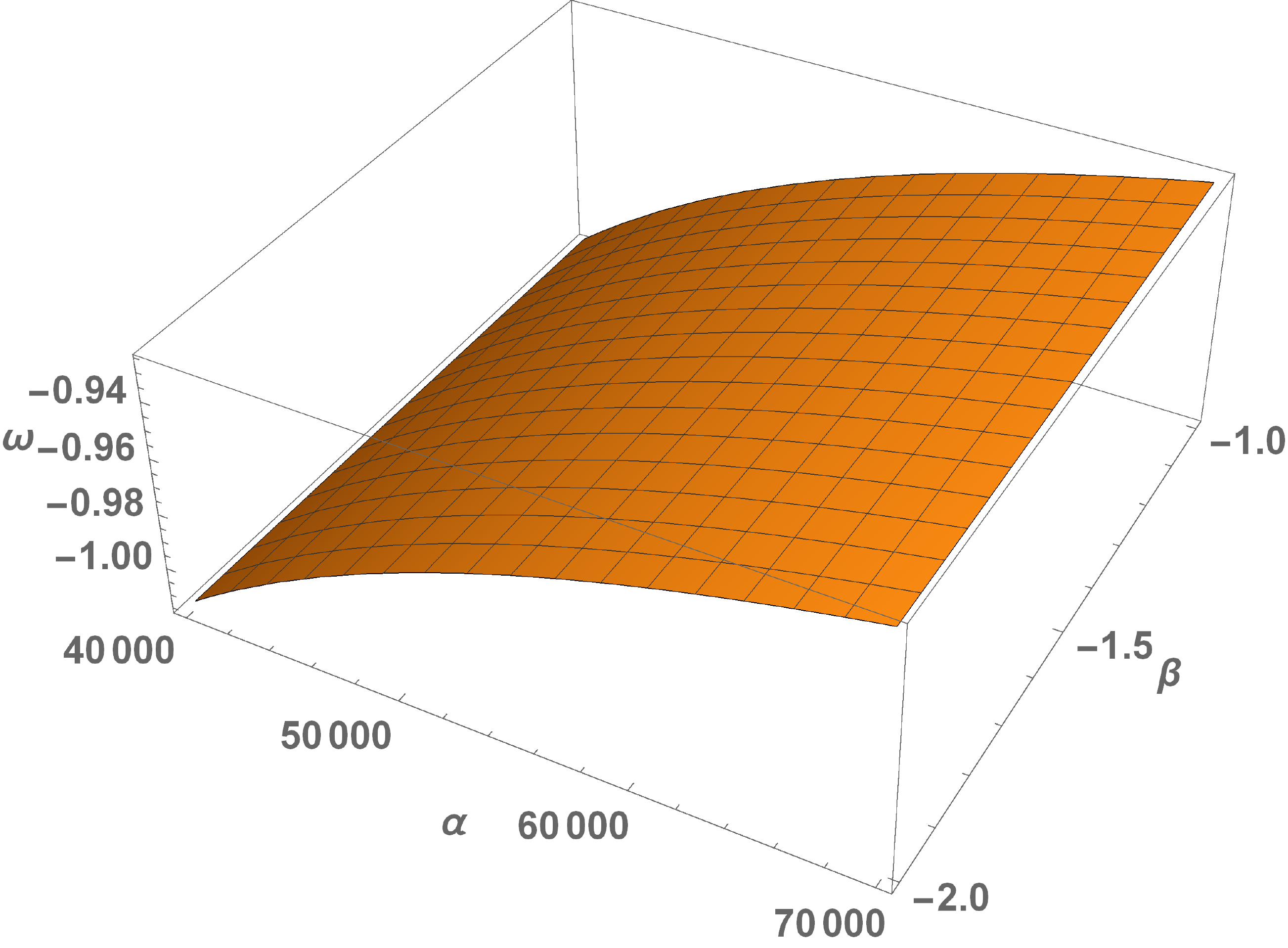}
         \caption{The EoS parameter.}
         \label{fig:omega2}
     \end{subfigure}
        \caption{The behavior of the density, pressure and the EoS parameters considering $F(R)= R+ \alpha log(\beta R)$ for Case II with $ 40000 \leq \alpha \leq 70000$ and $-2\leq \beta \leq -1$.}
        \label{fig:model1 Case2}
\end{figure}

We consider the present values $q_{0}= -0.55$, $j_{0}=1$, $s_{0}= -0.35$ and $H_{0}=67.9 $ km/s/Mpc to study the behavior of different parameters. It is observed from Figs. \ref{fig:density2} and \ref{fig:pressure2} that density is positive whereas pressure is negative in the range $ 40000 \leq \alpha \leq 70000$ and $-2\leq \beta \leq -1$. As a result, the EoS parameter is used to constrain model parameters to investigate various energy conditions.
In Fig. \ref{fig:omega2}, it is seen that the EoS parameter lies in the quintessence phase, and $\omega$ is close to -1, which shows an accelerated phase of the universe. 

\subsection{Model-II}

\textbf{Case 2.1.2 : $\dot{R} +3 R \neq 0$ } 

We assume the functional form of $F(R)= \gamma Exp(\frac{\zeta}{R})-R$ \cite{Girones/2010}, where $\gamma$ and $\zeta$ are constants. In the limiting case when $\gamma$ tends to $0$, the model reduces to the case of GR. We consider the present values of $q_{0}= -0.55$, $j_{0}=1$, $s_{0}= -0.35$ and $H_{0}=67.9 $ km/s/Mpc \cite{Capo/2019, Planck/2018}. Then the expressions for density $\rho$ and pressure $p$ read as,
\begin{equation}
\rho = \frac{\frac{1}{2} \gamma  e^{\frac{\zeta }{6 H^2 (q-1)}}-3 H^2 (q-1)-\frac{\gamma  \zeta  (-j+q+2) (H (-j+q+2)+4 (q-1)) e^{\frac{\zeta }{6 H^2 (q-1)}} \left(\zeta +12 H^2 (q-1)\right)}{144 H^5 (q-1)^4 (H (-j+q+2)+3 (q-1))}+\frac{(H (j-q-2)-2 q+2) \left(36 H^4 (q-1)^2+\gamma  \zeta  e^{\frac{\zeta }{6 H^2 (q-1)}}\right)}{24 H^2 (q-1)^2}}{\kappa}
\end{equation}

\begin{multline}
p= \frac{1}{\kappa} \left(-\frac{1}{2} \gamma  e^{\frac{\zeta }{6 H^2 (q-1)}}+3 H^2 (q-1)+\frac{\splitfrac{\gamma  \zeta  e^{\frac{\zeta }{6 H^2 (q-1)}} \left(\zeta +12 H^2 (q-1)\right) (H^2 (-(j-q-2)) (q (q+8)+s+6)+2 H (7 j^2}{-14 j (q+2)+q (q (6 q+49)+6 s+16)-6 s-8)+8 (-j q+j+q^2+q-2))}}{864 H^5 (q-1)^4 (H (-j+q+2)+3 (q-1))}\right.\\
\left.+\frac{(H (j-q-2)-6 q+6) \left(36 H^4 (q-1)^2+\gamma  \zeta  e^{\frac{\zeta }{6 H^2 (q-1)}}\right)}{72 H^2 (q-1)^2}+\frac{\gamma  \zeta  (-j+q+2)^2 e^{\frac{\zeta }{6 H^2 (q-1)}} \left(\zeta ^2+216 H^4 (q-1)^2+36 \zeta  H^2 (q-1)\right)}{1296 H^6 (q-1)^6}\right)
\end{multline}

\begin{figure}[H]
     \centering
     \begin{subfigure}[b]{0.45\textwidth}
         \centering
         \includegraphics[width=\textwidth]{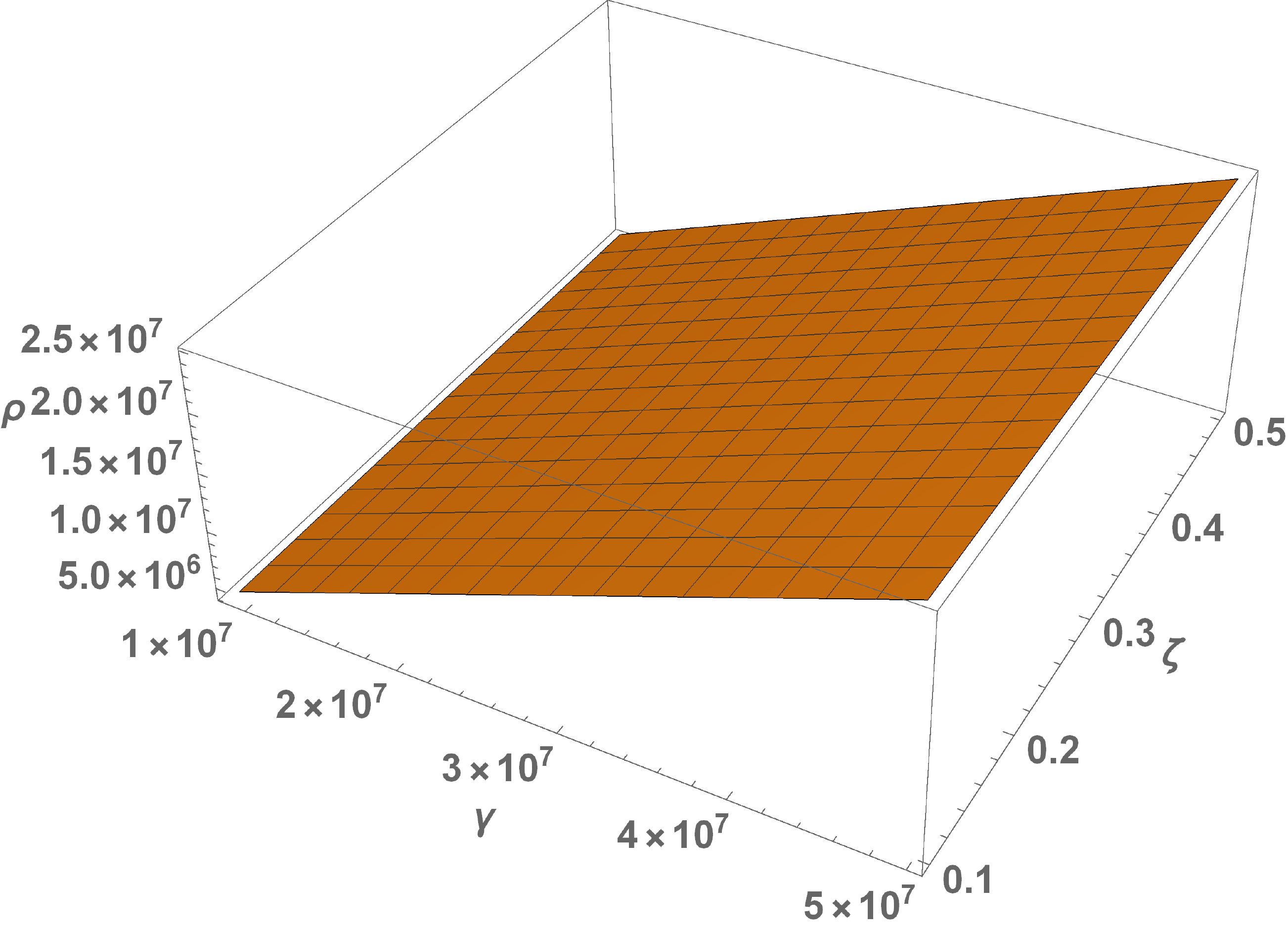}
         \caption{The density parameter.}
         \label{fig:density model2}
     \end{subfigure}
     \hfill
     \begin{subfigure}[b]{0.45\textwidth}
         \centering
         \includegraphics[width=\textwidth]{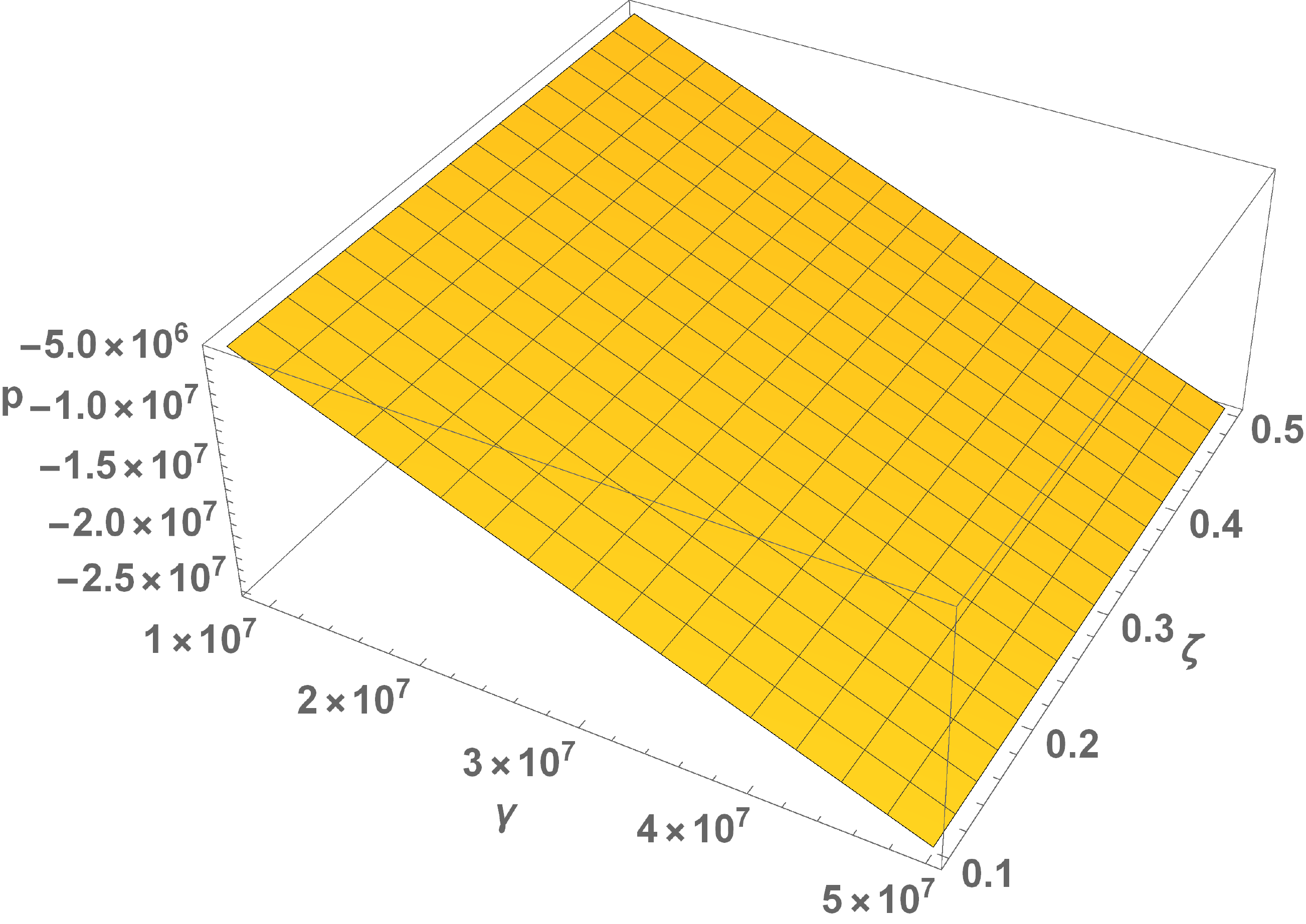}
         \caption{The pressure parameter.}
         \label{fig:pressure model2}
     \end{subfigure}
     \hfill
     \begin{subfigure}[b]{0.45\textwidth}
         \centering
         \includegraphics[width=\textwidth]{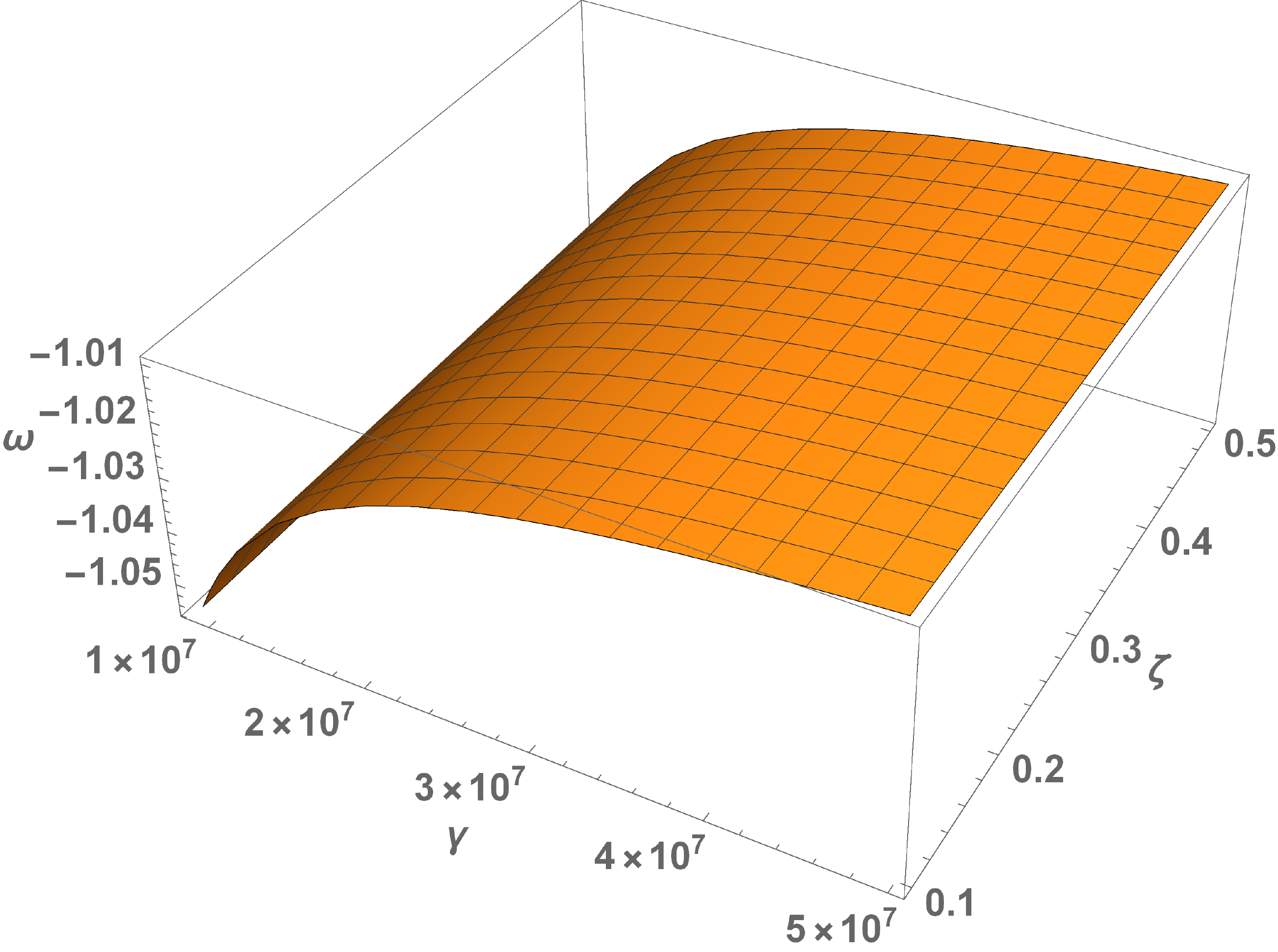}
         \caption{The EoS parameter.}
         \label{fig:omega model2}
     \end{subfigure}
        \caption{The behavior of the density, pressure and the EoS parameters considering $F(R)= \gamma Exp(\zeta/R) -R$ for Case 2.1.2 with $ 9 \times 10^{6} \leq \gamma \leq 5 \times 10^{7}$ and $0.1 \leq \zeta \leq 0.5$.}
        \label{fig:model2 Case1}
\end{figure}

We obtained the bound for $\gamma$ and $\zeta$ through the density parameter in \ref{fig:density model2}, pressure \ref{fig:pressure model2} and the EoS parameter in \ref{fig:omega model2}. According to the observations, the density behaves positively, whereas the EoS parameter should be negative and close to -1. 
We notice that $\omega \approx -1$ from lower values in, revealing an unexpected behavior consistent with a phantom era for dark energy. Furthermore, when dealing with a standard explanation of gravity plus an essential phantom field, a phantom period for dark energy could lead to a problematic description for the universe because the DEC would be violated \cite{Carroll/2005}. Moreover, the equation of state parameter is bound to be greater than -1, implying that DEC is validated \cite{Carroll/2003}. However, given the existing experimental bounds for $\omega = -1.03^{+0.03}_{-0.03}$ \cite{Planck/2018} show that a phantom description of dark energy is still possible.\\

\textbf{Case II : $\dot{R} = 0$ } 

In this case $R$ is constant. We obtain $\rho$ and $p$ for this case in section \ref{sec2}. Using equations \eqref{rho2}, \eqref{p2}, \eqref{R} and the functional form $F(R)= \gamma Exp(\frac{\zeta}{R})-R$, where $\gamma$, $\zeta$ are constants. So, the density and pressure are as follows.

\begin{equation}
\rho = \frac{6 H^2 (q-1) \left(\gamma  \zeta  e^{6 \zeta  H^2 (q-1)}-2\right)+\gamma  e^{6 \zeta  H^2 (q-1)}}{2 \kappa}.
\end{equation} 

\begin{equation}
p=\frac{\gamma  e^{6 \zeta  H^2 (q-1)} \left(6 \zeta  H^2 (q-1)-1\right)}{2 \kappa}.
\label{p}
\end{equation}

The Eos parameter $\omega=\frac{p}{\rho}$ is obtained as 

\begin{equation}
\omega= \frac{\gamma  e^{6 \zeta  H^2 (q-1)} \left(6 \zeta  H^2 (q-1)-1\right)}{6 H^2 (q-1) \left(\gamma  \zeta  e^{6 \zeta  H^2 (q-1)}-2\right)+\gamma  e^{6 \zeta  H^2 (q-1)}}.
\end{equation}

\begin{figure}[H]
     \centering
     \begin{subfigure}[b]{0.45\textwidth}
         \centering
         \includegraphics[width=\textwidth]{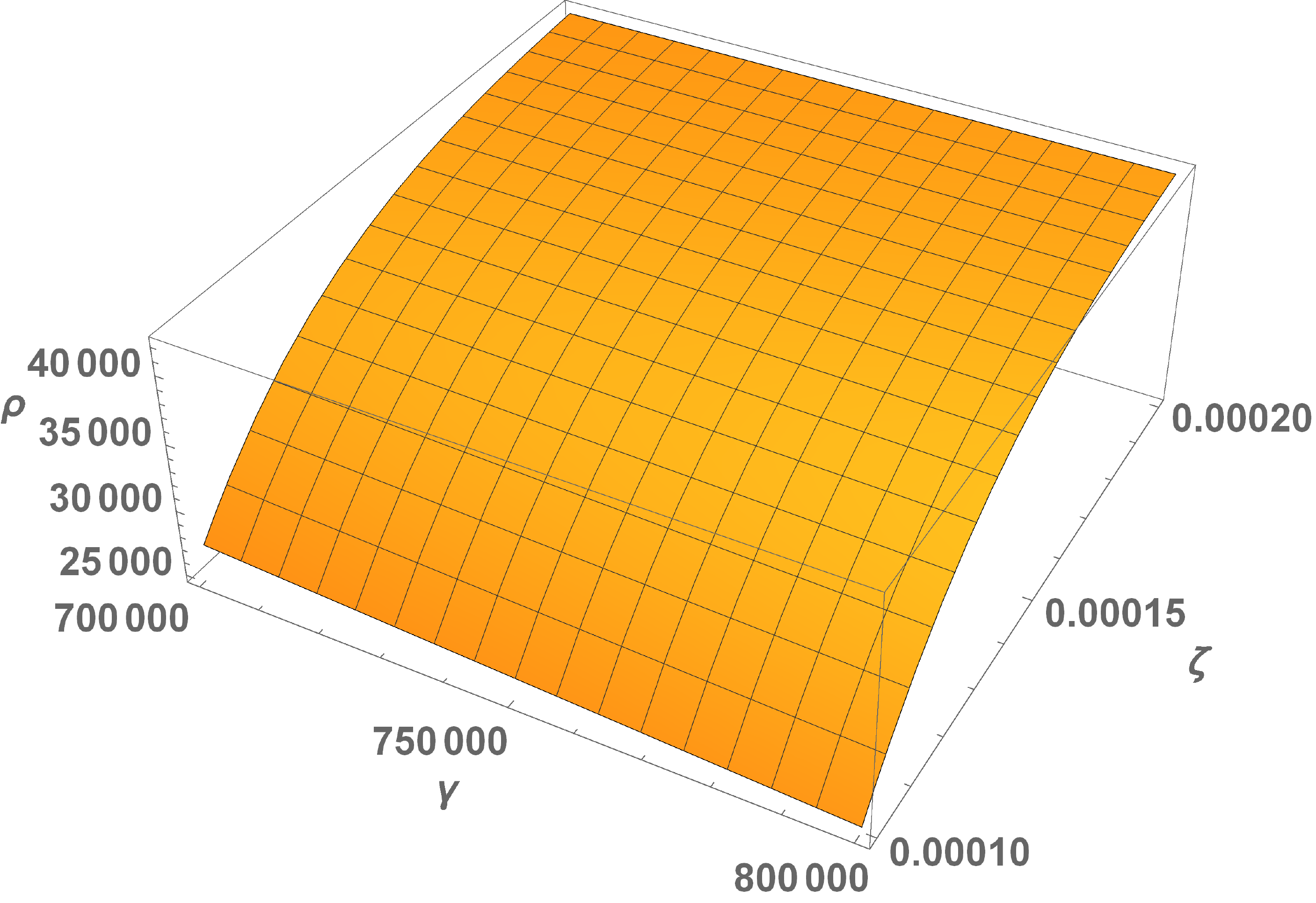}
         \caption{The density parameter.}
         \label{fig:density case2 model2}
     \end{subfigure}
     \hfill
     \begin{subfigure}[b]{0.45\textwidth}
         \centering
         \includegraphics[width=\textwidth]{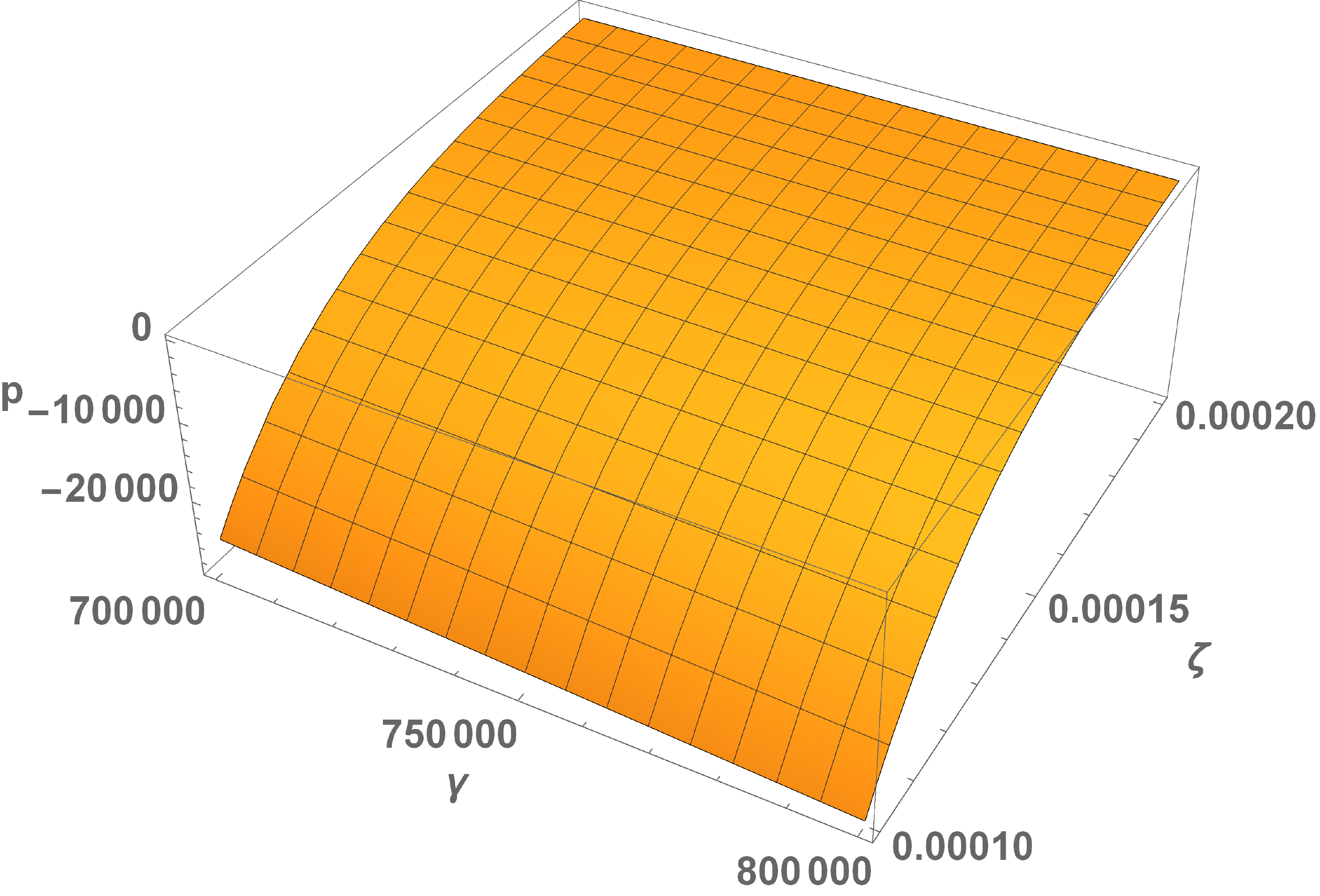}
         \caption{The pressure parameter.}
         \label{fig:pressure case2 model2}
     \end{subfigure}
     \hfill
     \begin{subfigure}[b]{0.45\textwidth}
         \centering
         \includegraphics[width=\textwidth]{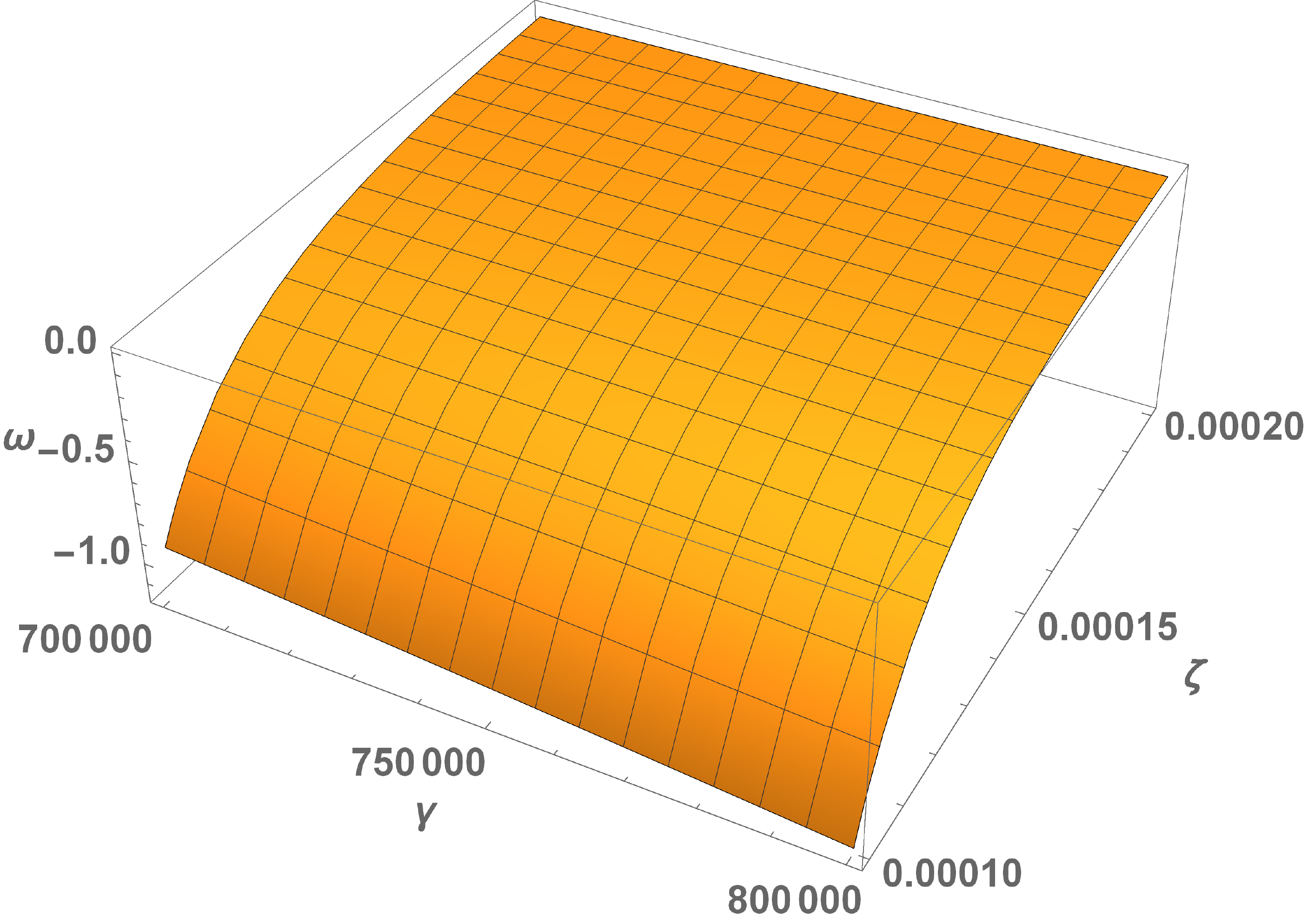}
         \caption{The EoS parameter.}
         \label{fig:omega case2 model2}
     \end{subfigure}
        \caption{The behavior of the density, pressure and the EoS parameters considering $F(R)= \gamma Exp(\frac{\zeta}{R})-R$ for Case II with $ 7 \times 10^{5} \leq \gamma \leq 8 \times 10^{5}$ and $0.0001 \leq \zeta \leq 0.0002$.}
        \label{fig:model2 Case2}
\end{figure}

We consider the present values $q_{0}= -0.55$, $j_{0}=1$, $s_{0}= -0.35$ and $H_{0}=67.9 $ km/s/Mpc to study the behavior of different parameters. It is observed from Figs. \ref{fig:density case2 model2} and \ref{fig:pressure case2 model2} that density is positive whereas pressure is negative in the range $ 7 \times 10^{5} \leq \gamma \leq 8 \times 10^{5}$ and $0.0001 \leq \zeta \leq 0.0002$. Here, we have considered the bounds of parameters large due to higher powers of $H$.
In Fig. \ref{fig:omega case2 model2}, it is seen that the EoS parameter lies in the quintessence phase, with a transition from positive to negative and is close to -1, depicting an accelerated phase of the universe. 

\section{Energy conditions} \label{sec4}

Raychaudhuri equations, which play a fundamental part in any consideration of the congruence of null and timelike geodesics with the criterion that only gravity is attractive but also the energy density is positive, give rise to the concept of energy conditions. Let $u^{i}$ be the tangent vector field to a congruence of timelike geodesics in a spacetime manifold furnished with a  metric $g_{ij}$ to illustrate this concept.  The Raychaudhuri equations describe the temporal evolution of the expansion scalar. For a brief review see\cite{Sayan/2007}.

\begin{equation}
\frac{d \theta}{d \tau}= -\frac{\theta^{2}}{3} + \omega_{ij} \omega^{ij} - \sigma_{ij} \sigma^{ij} - R_{ij} u^{i} u^{j},
\end{equation}

\begin{equation}
\frac{d \theta}{d \tau}= -\frac{\theta^{2}}{2} + \omega_{ij} \omega^{ij} - \sigma_{ij} \sigma^{ij} - R_{ij} k^{i} k^{j}.
\end{equation}

where $\omega_{ij}$, $\sigma_{ij}$, $u^{i}$ and $k^{i}$ are the rotation, shear tensor, timelike and null tangent vector fields respectively. Since, we are considering the case $\omega_{ij}=0$ and neglecting the small distortions, we obtained $\theta= -\tau R_{ij} u^{i} u^{j}$ and $\theta= -\tau R_{ij} k^{i} k^{j}$. According to attractive nature of gravity, $R_{ij} u^{i} u^{j} \geq 0$ and $R_{ij} k^{i} k^{j} \geq 0$. Further according to Einstein field equations, we get $(T_{ij}-\frac{1}{2} g_{ij} T)u^{i} u^{j} \geq 0$ and $(T_{ij}-\frac{1}{2} g_{ij} T)k^{i} k^{j} \geq 0$. 
In case of perfect fluid matter distribution, the energy conditions are defined as 
\begin{itemize}
\item Null energy condition (NEC): $\rho + p \geq 0$ 
\item Weak energy condition (WEC): $\rho + p \geq 0$ and $\rho \geq 0$ 
\item Strong energy conditions (SEC): $\rho + p \geq 0$ and $\rho + 3 p \geq 0$ 
\item Dominant energy condition (DEC): $\rho \pm p \geq 0$ and $\rho \geq 0$.
\end{itemize}

The concept of energy boundaries in modified theories of gravity can be expanded with the premise that the whole cosmic matter behaves like a perfect fluid due to the utterly geometric nature of the Raychaudhuri equations. So, we can consider $p^{eff}$ as $p$ and $\rho^{eff}$ as $\rho$. 
The classical energy conditions and their cosmological implications were studied extensively in $F(R)$-gravity theory in FLRW setting \cite{Santos/2007,Wang/2010}.

\subsection{Model I}

\textbf{Case 2.1.2: $\dot{R}+3R \neq 0$}

The energy conditions for the functional form of $F(R)= R+ \alpha log(\beta R)$ are the following.

\begin{multline}
\displaystyle \rho + p= \frac{H^2 }{2 \kappa}  \left(\frac{\splitfrac{3 F_{RR} H (H^2 (j-q-2) (q (q+8)+s+6)-4 H (2 j^2-4 j (q+2)+q (q (3 q+23)+3 s+2)-3 s-10)-}{16 (q-1) (j-q-2))}}{H (j-q-2)-3 q+3}+ \right.\\ 
\left.  4 F_{R} (H (-j+q+2)+3 (q-1))-72 F_{RRR} H^4 (-j+q+2)^2\right)
\end{multline}

\begin{multline}
\displaystyle \rho - p = \frac{1}{2 \kappa} \left(\frac{\splitfrac{3 F_{RR} (H^2 (-(j-q-2)) (q (q+8)+s+6)+4 H (5 j^2-10 j (q+2)+q (q (3 q+26)+3 s+14)-3 s+2)-}{32 (q-1) (j-q-2))}}{H (j-q-2)-3 q+3} +\right.\\
\left. 2 F + H^3(2 F_{R} (-j+q+2)+72 F_{RRR} H^3 (-j+q+2)^2 \right)
\end{multline}

\begin{multline}
\displaystyle \rho + 3p = \frac{1}{2 \kappa}  \left(\frac{9 F_{RR} H^4 (H (j-q-2) (q (q+8)+s+6)-12 (j^2-2 j (q+2)+q (q (q+8)+s+2)-s-2))}{H (j-q-2)-3 q+3} \right.\\
\left. -2 F+6 F_{R} H^2 (H (-j+q+2)+4 (q-1))-216 F_{RRR} H^6 (-j+q+2)^2 \right)
\end{multline}

As, WEC ( $\rho \geq 0$ ) depends on the observational values of $q_{0}$ and $j_{0}$, i.e. independent of $s_{0}$, so we obtain the restrictions on the model parameters $\alpha$ and $\beta$. Since no accurate measurement of the snap parameter has been reported, we can concentrate on the WEC requirement in confronting energy conditions. The condition reduces to 
\begin{equation}
0.5 \alpha  \log (-42876.8 \beta )-4.42758 \alpha +168430 \geq 0.
\end{equation}  
It is seen that $\beta$ can take negative values and $\alpha$ can vary negative to positive satisfying the above condition.

\begin{figure}[H]
     \centering
     \begin{subfigure}[b]{0.45\textwidth}
         \centering
         \includegraphics[width=\textwidth]{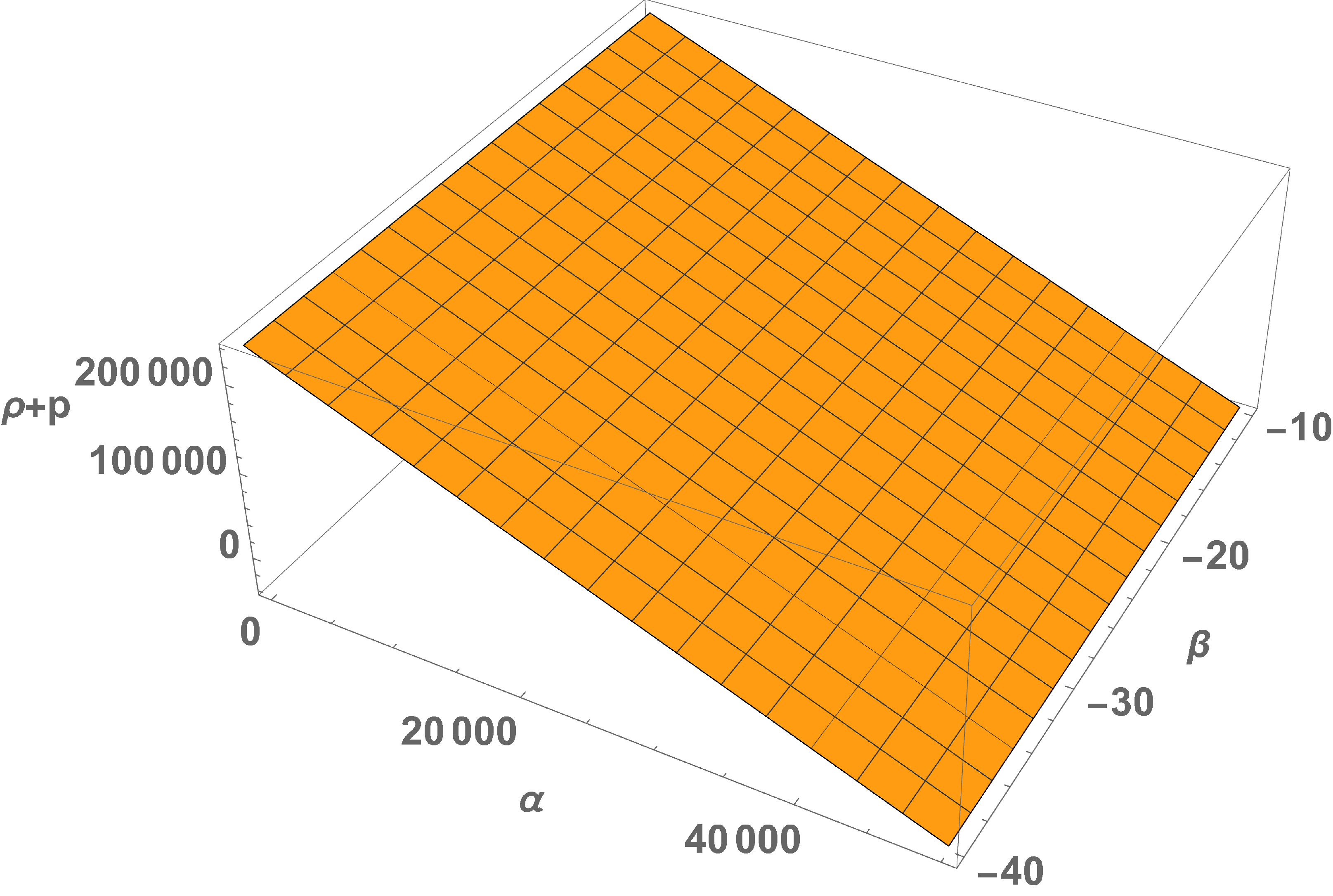}
         \caption{$\rho$ + p.}
         \label{fig:NEC}
     \end{subfigure}
     \hfill
     \begin{subfigure}[b]{0.45\textwidth}
         \centering
         \includegraphics[width=\textwidth]{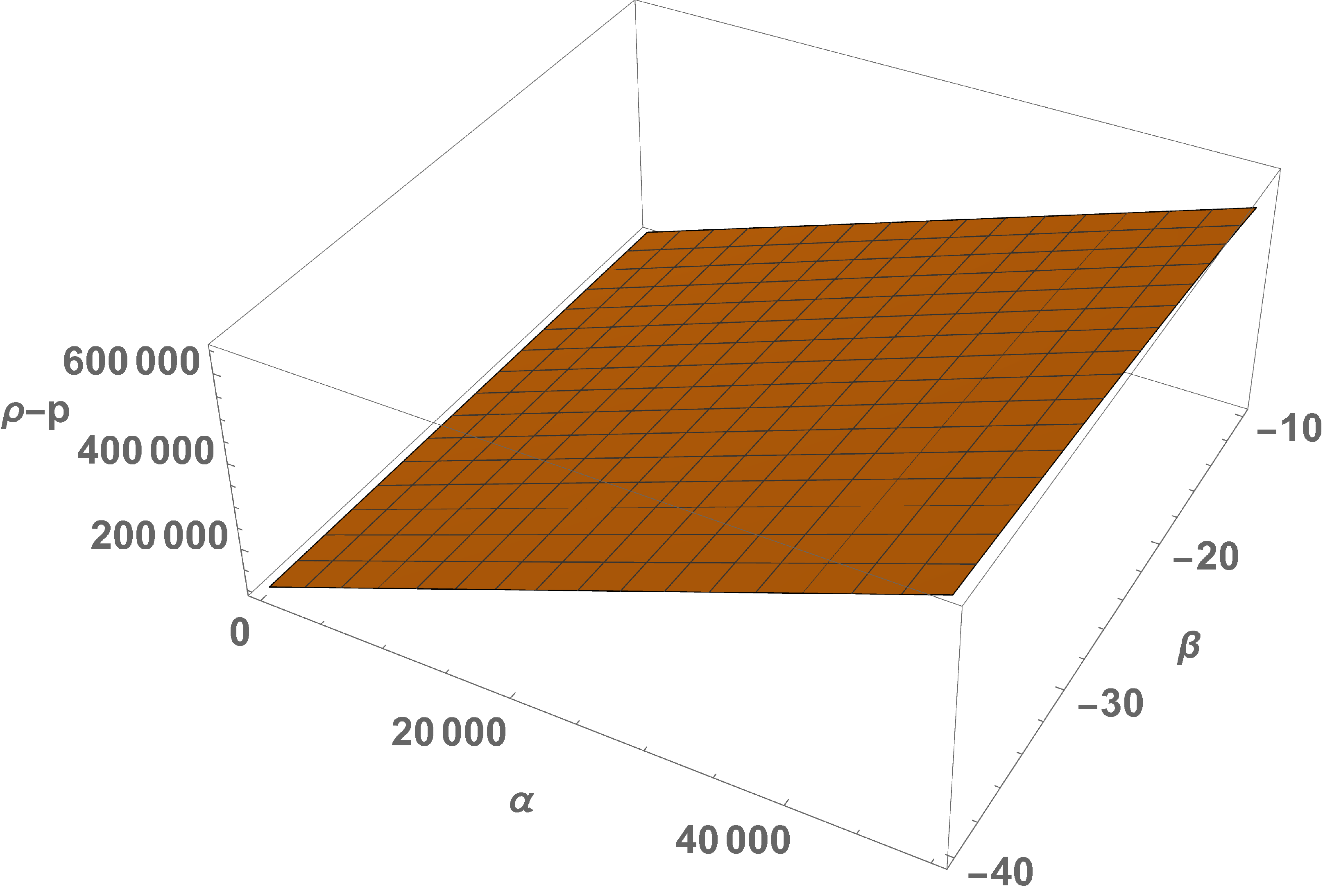}
         \caption{$\rho$ - p.}
         \label{fig:DEC}
     \end{subfigure}
     \hfill
     \begin{subfigure}[b]{0.45\textwidth}
         \centering
         \includegraphics[width=\textwidth]{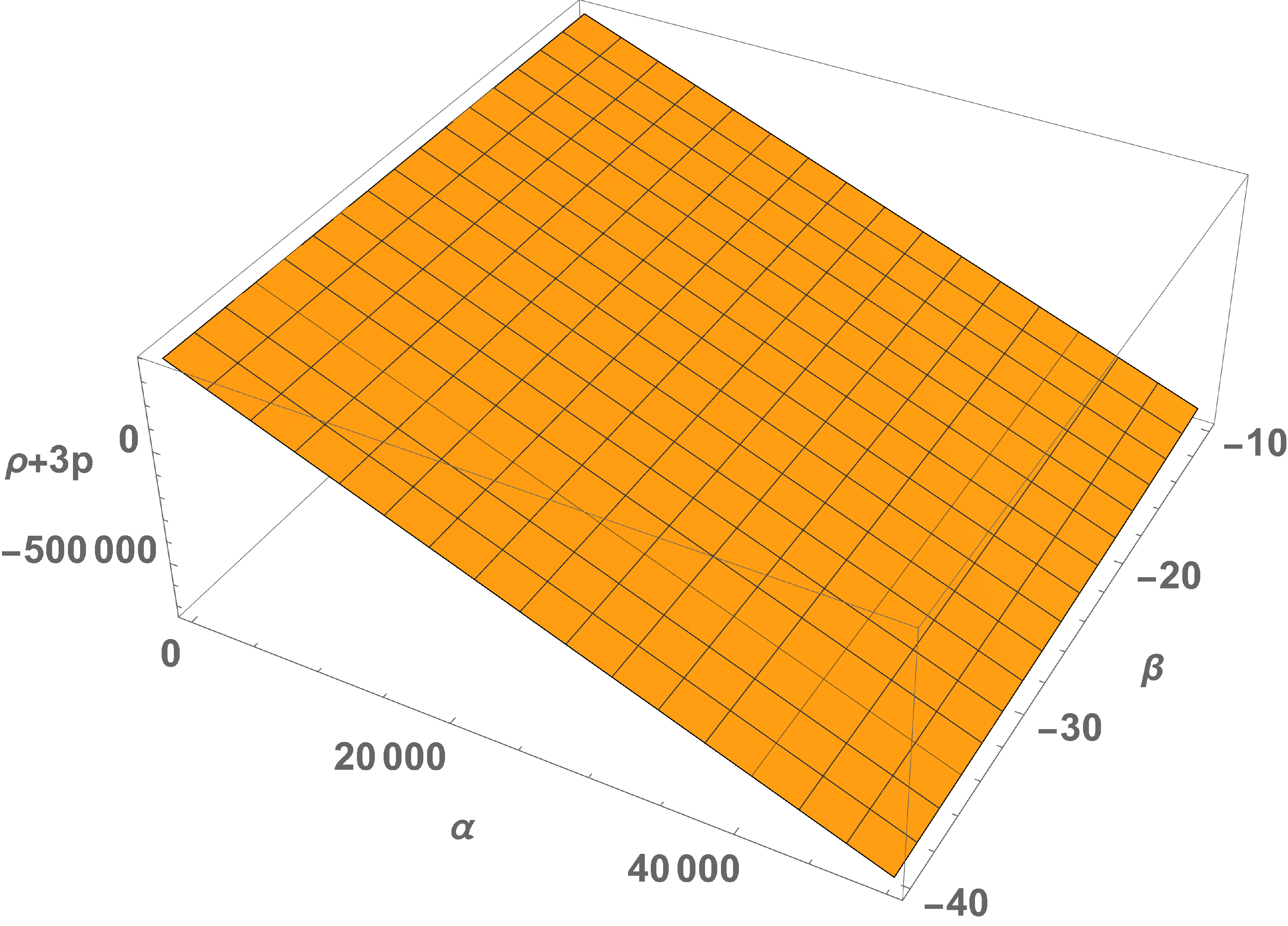}
         \caption{$\rho$ + 3 p.}
         \label{fig:SEC}
     \end{subfigure}
        \caption{The behavior of energy conditions considering $F(R)= R+ \alpha log(\beta R)$ for Case 2.1.2 with $ -1 \leq \alpha \leq 50000$ and $-40\leq \beta \leq -10$.}
        \label{fig:EC model1}
\end{figure}

There may be a violation of energy conditions in some instances without the system becoming unacceptable. The success of the hypothesis of inflation and recent observations of cosmic acceleration point to a violation of SEC in the universe. So, we can observe from the plots \ref{fig:NEC}, \ref{fig:DEC} that the WEC, DEC are satisfying their conditions (showing positive behavior) in the range of $ -1 \leq \alpha \leq 50000$ and $-40\leq \beta \leq -10$. On the other, SEC in Fig. \ref{fig:SEC} exhibits  negative behavior violating its condition and hence supports an accelerated expansion of the universe.\\

\textbf{Case II: $\dot{R}=0$}

The energy conditions for this case should satisfy the following:

\begin{equation}
\rho +p = \frac{\alpha +\alpha  \log \left(6 \beta  H^2 (q-1)\right)+12 H^2 (q-1)}{2 \kappa}+\frac{\alpha -\alpha  \log \left(6 \beta  H^2 (q-1)\right)}{2 \kappa} \geq 0
\end{equation}

\begin{equation}
\rho-p =\frac{\alpha +\alpha  \log \left(6 \beta  H^2 (q-1)\right)+12 H^2 (q-1)}{2 \kappa}-\frac{\alpha -\alpha  \log \left(6 \beta  H^2 (q-1)\right)}{2 \kappa} \geq 0
\end{equation}

\begin{equation}
 \rho+3p =\frac{\alpha +\alpha  \log \left(6 \beta  H^2 (q-1)\right)+12 H^2 (q-1)}{2 \kappa}-\frac{\alpha -\alpha  \log \left(6 \beta  H^2 (q-1)\right)}{2 \kappa} \geq 0 
\end{equation}

\begin{figure}[H]
     \centering
     \begin{subfigure}[b]{0.45\textwidth}
         \centering
         \includegraphics[width=\textwidth]{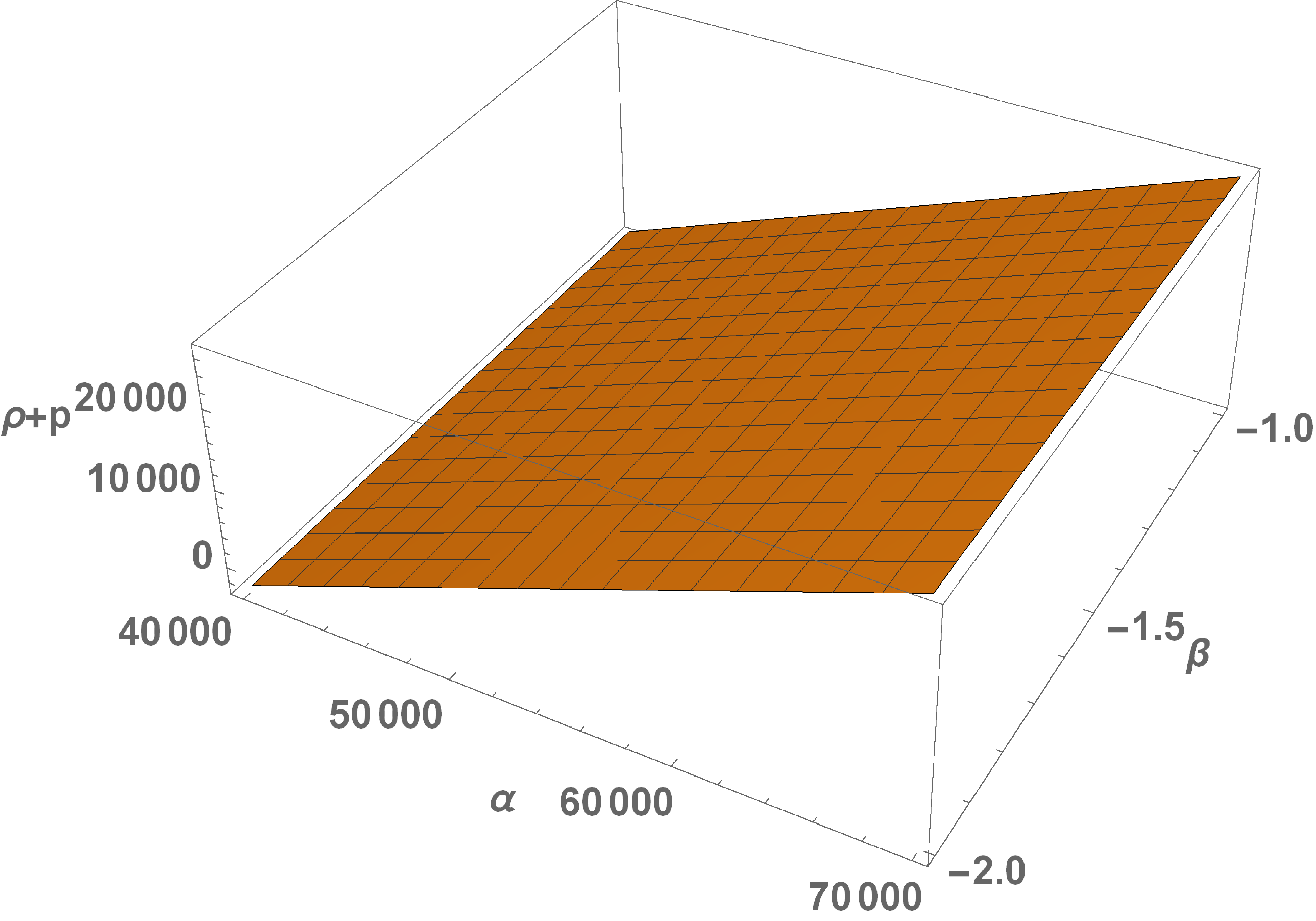}
         \caption{$\rho$ + p.}
         \label{fig:NEC2}
     \end{subfigure}
     \hfill
     \begin{subfigure}[b]{0.45\textwidth}
         \centering
         \includegraphics[width=\textwidth]{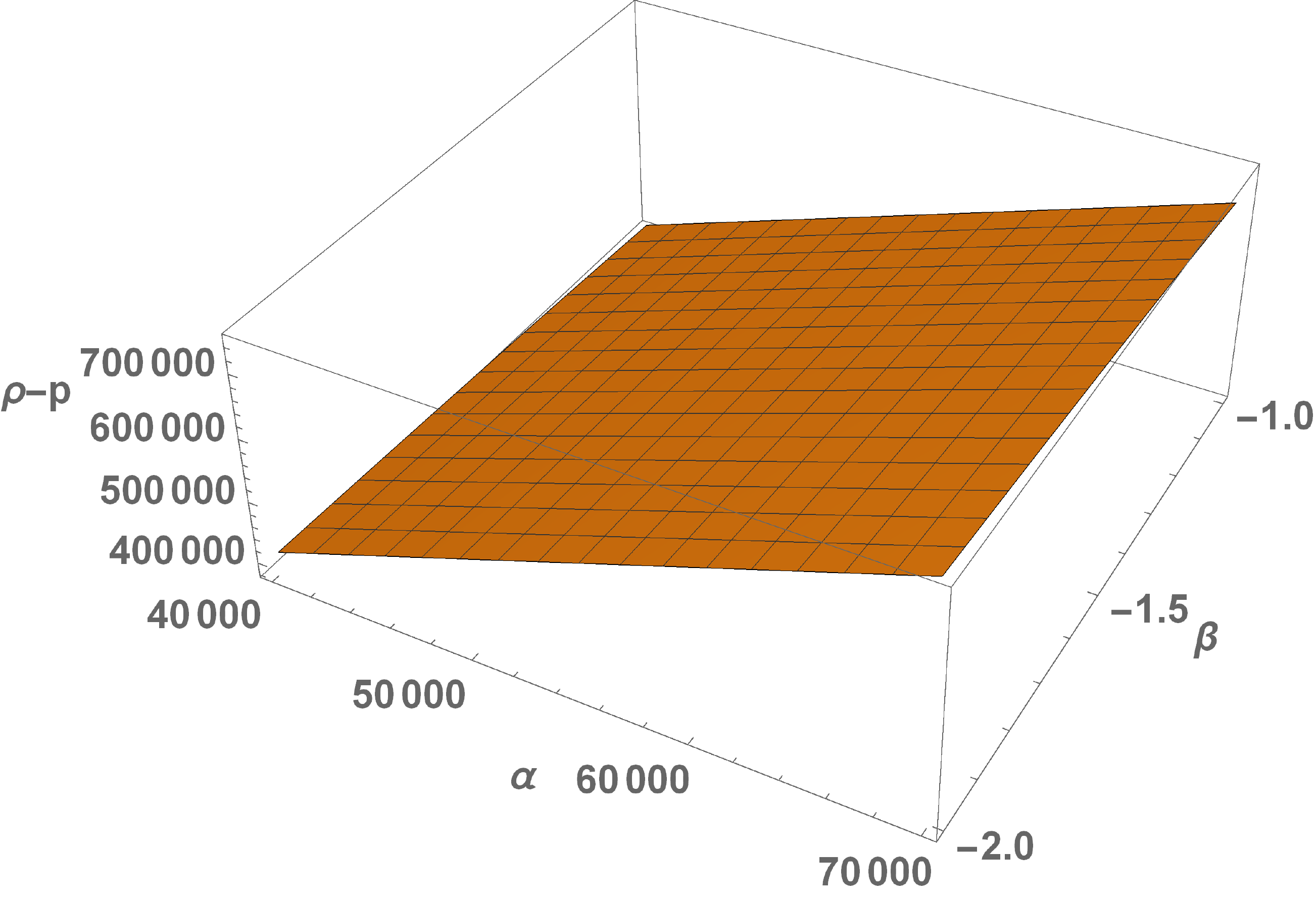}
         \caption{$\rho$ - p.}
         \label{fig:DEC2}
     \end{subfigure}
     \hfill
     \begin{subfigure}[b]{0.45\textwidth}
         \centering
         \includegraphics[width=\textwidth]{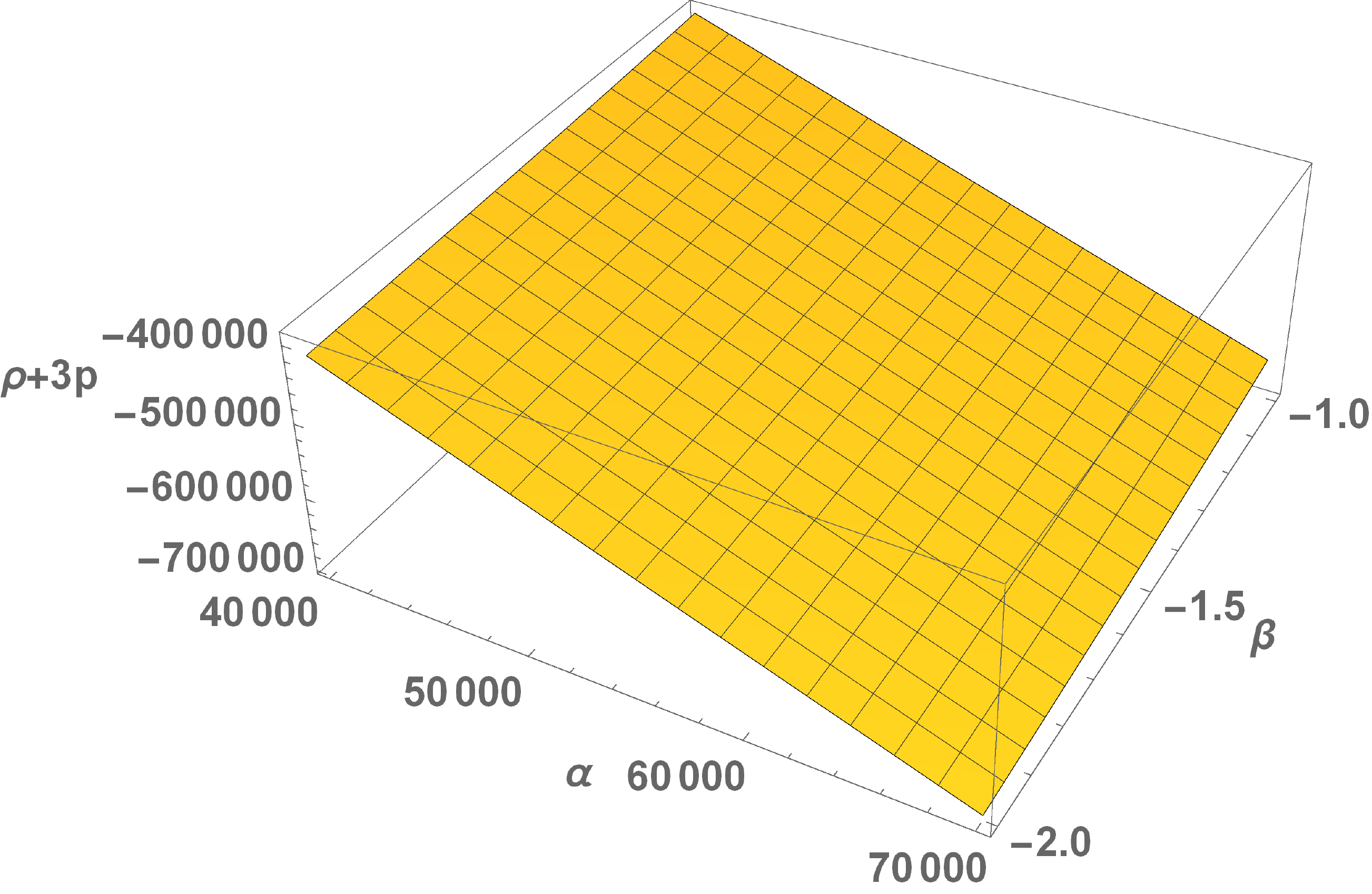}
         \caption{$\rho$ + 3 p.}
         \label{fig:SEC2}
     \end{subfigure}
        \caption{The behavior of energy conditions considering  $F(R)= R+ \alpha log(\beta R)$ for Case II with $ 40000 \leq \alpha \leq 70000$ and $-2\leq \beta \leq -1$ .}
        \label{fig:EC model1 case2}
\end{figure}

WEC is the combination of $\rho+p \geq 0$ and the positive density. The condition on WEC gives 
\begin{equation}
0.5 (\alpha  \log (-42876.8 \beta )+\alpha -85753.6) \geq 0.
\end{equation}
which confronts the negative value of $\beta$ and the variation of $\alpha$.
Here we can observe from Figs. \ref{fig:NEC2} \& \ref{fig:DEC2} that WEC, DEC  are validating their respective conditions with $ 40000 \leq \alpha \leq 70000$ and $-2\leq \beta \leq -1$.  On the other hand, SEC in Fig. \ref{fig:SEC2} shows negative behavior violating its condition in the same range. Hence, the violation of the SEC supports an accelerated expansion of the universe.

\subsection{Model-II}

\textbf{Case 2.1.2: $\dot{R}+3R \neq 0$}\\

The energy conditions for the functional form $F(R)= \gamma Exp(\frac{\zeta}{R})-R$ are obtained and the behavior is shown in the following plots.

\begin{figure}[H]
     \centering
     \begin{subfigure}[b]{0.45\textwidth}
         \centering
         \includegraphics[width=\textwidth]{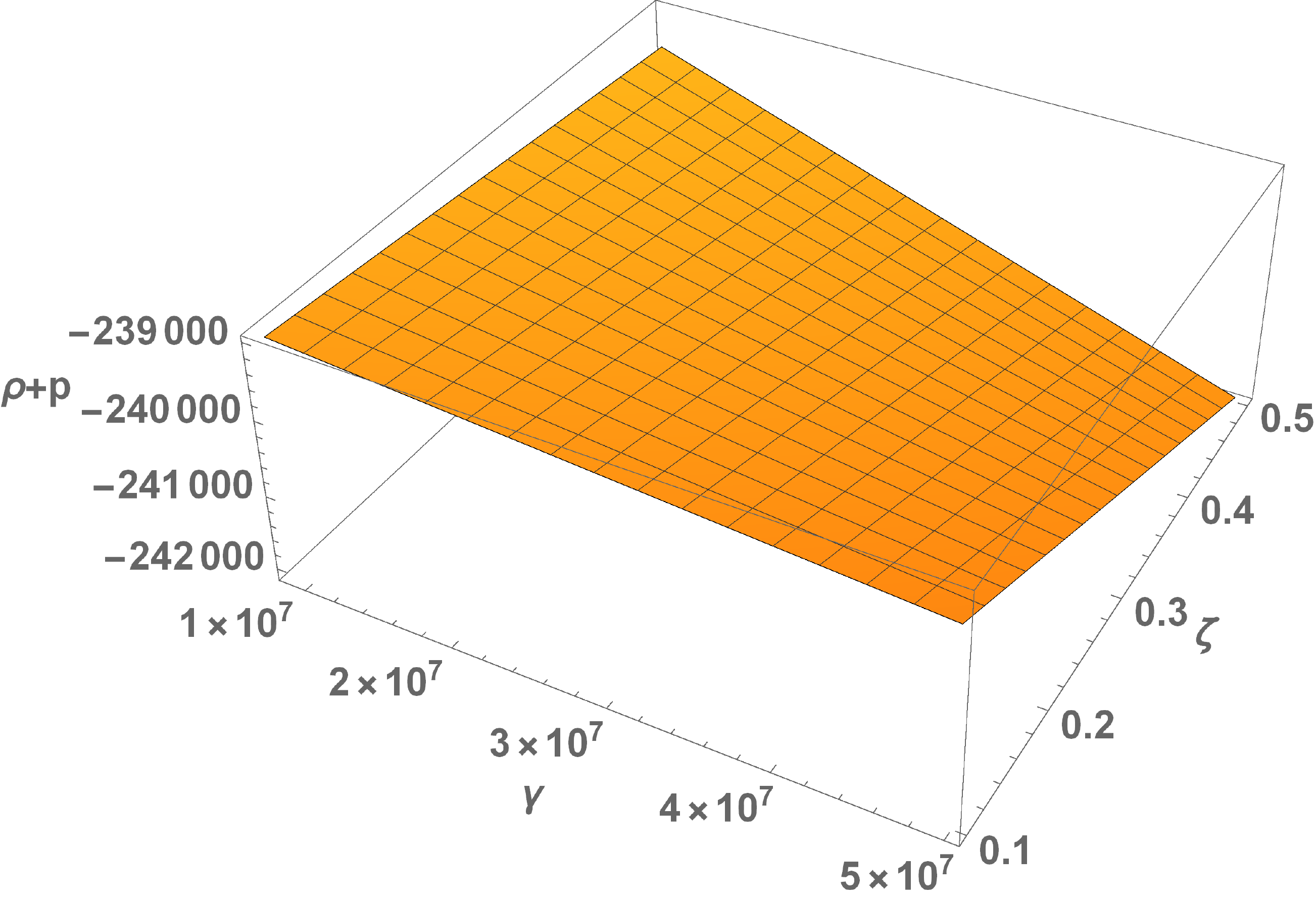}
         \caption{$\rho$ + p.}
         \label{fig:NEC model2}
     \end{subfigure}
     \hfill
     \begin{subfigure}[b]{0.45\textwidth}
         \centering
         \includegraphics[width=\textwidth]{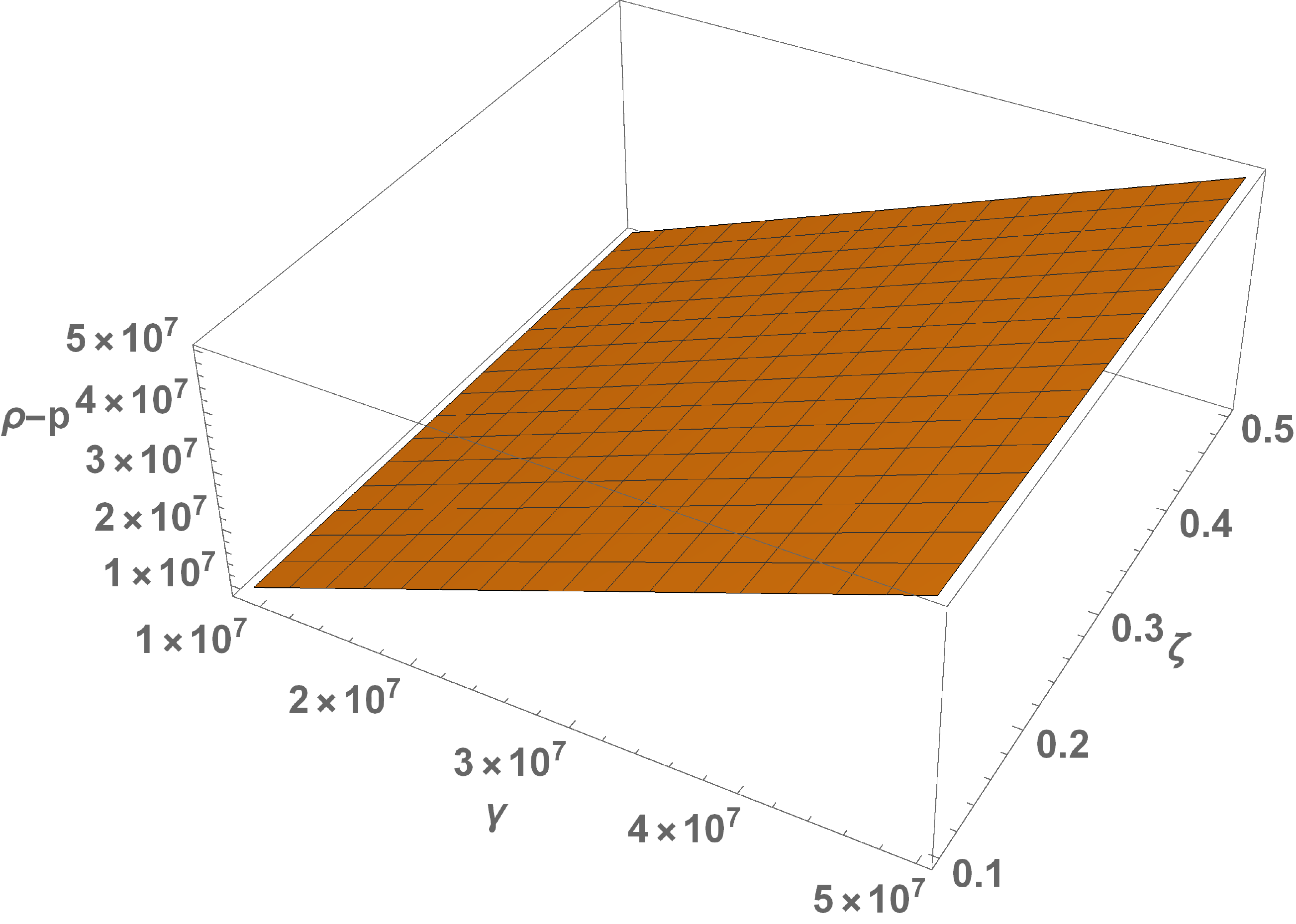}
         \caption{$\rho$ - p.}
         \label{fig:DEC model2}
     \end{subfigure}
     \hfill
     \begin{subfigure}[b]{0.45\textwidth}
         \centering
         \includegraphics[width=\textwidth]{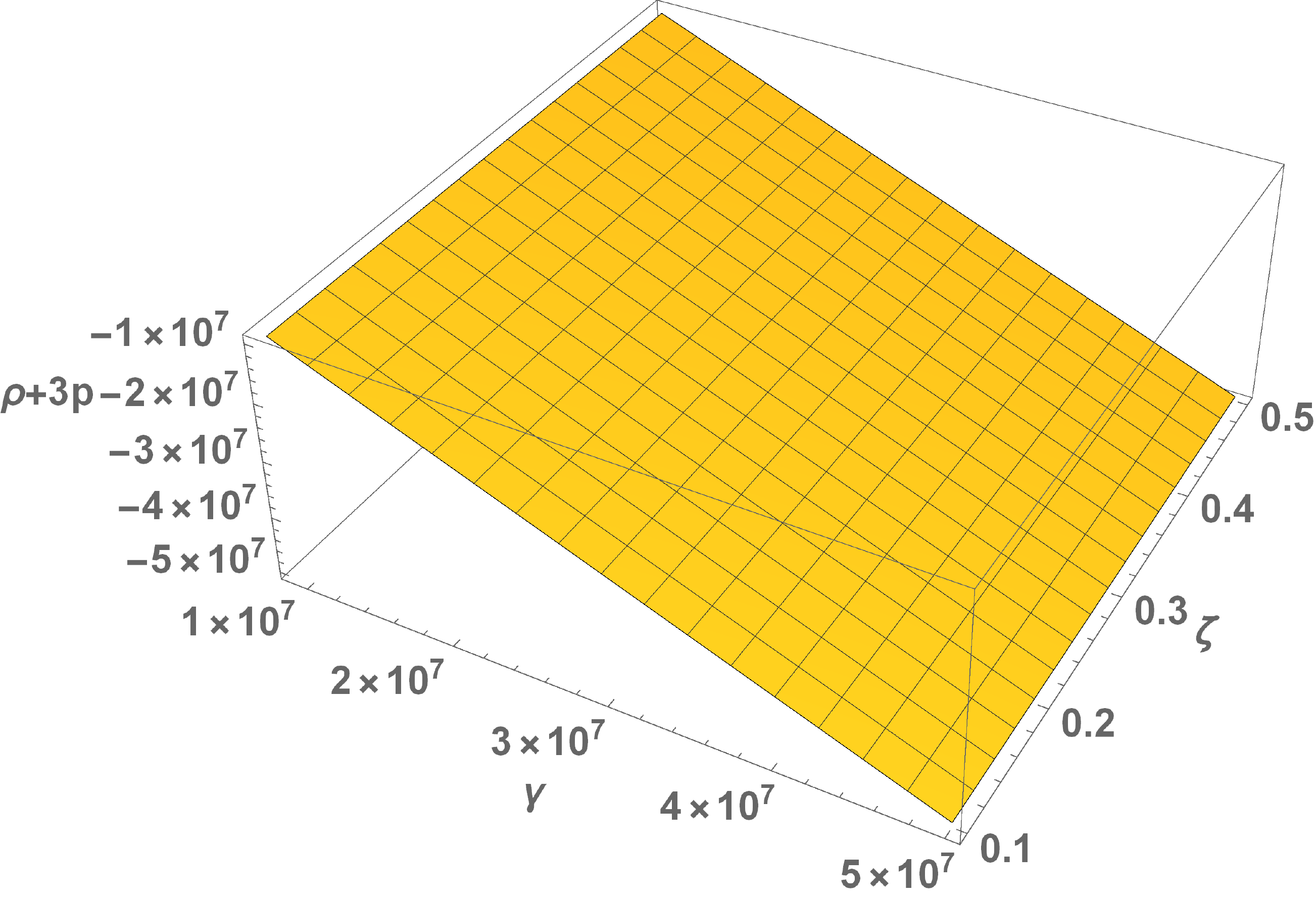}
         \caption{$\rho$ + 3 p.}
         \label{fig:SEC model2}
     \end{subfigure}
        \caption{The behavior of energy conditions considering $F(R)= \gamma Exp(\zeta/R) -R$ for Case 2.1.2 with $ 9 \times 10^{6} \leq \gamma \leq 5 \times 10^{7}$ and $0.1 \leq \zeta \leq 0.5$.}
        \label{fig:EC model2}
\end{figure}

Furthermore, the plot \ref{fig:EC model2} reveals that DEC in \ref{fig:DEC model2} is satisfying its conditions whereas WEC in \ref{fig:NEC model2} is partially obeyed(i.e. satisfying only $\rho > 0$). However, we can also see that SEC in \ref{fig:SEC model2} is again violated confirming that our universe experiences an accelerated expansion of the universe. Hence, the WEC violation along with the positive density acts as a scalar-tensor gravity model in this case \cite{Whinnett/2004}, and such a violation can be understood as natural contributions from quantum processes to classical gravity \cite{Calcagni/2017}.\\ 

\textbf{Case II: $\dot{R} =0$}\\

The energy conditions for the functional form $F(R)= \gamma Exp(\frac{\zeta}{R})-R$ are obtained. Since equations are little big. We have shown the behavior in the following plots.

\begin{figure}[H]
     \centering
     \begin{subfigure}[b]{0.45\textwidth}
         \centering
         \includegraphics[width=\textwidth]{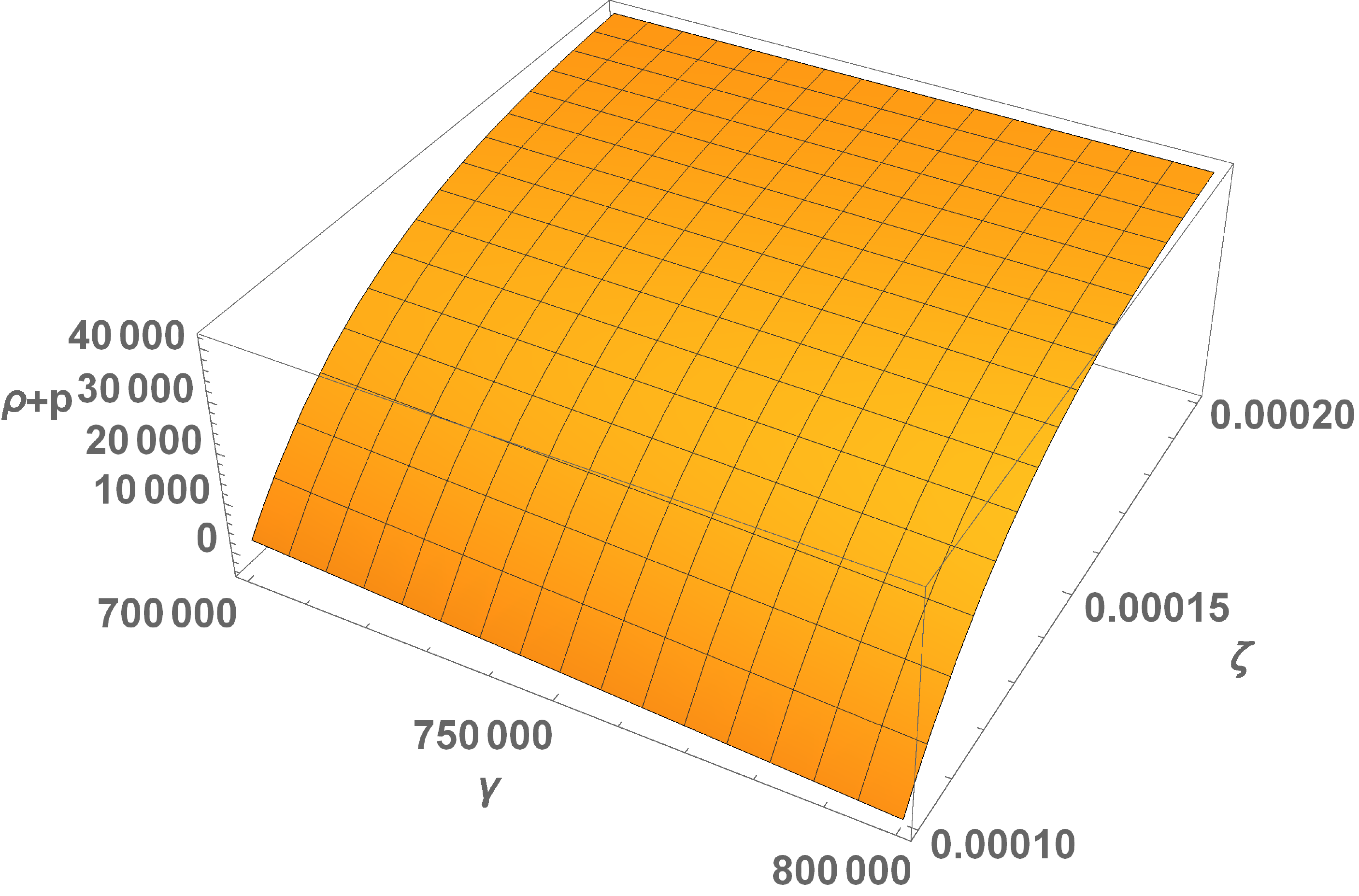}
         \caption{$\rho$ + p.}
         \label{fig:NEC case2 model2}
     \end{subfigure}
     \hfill
     \begin{subfigure}[b]{0.45\textwidth}
         \centering
         \includegraphics[width=\textwidth]{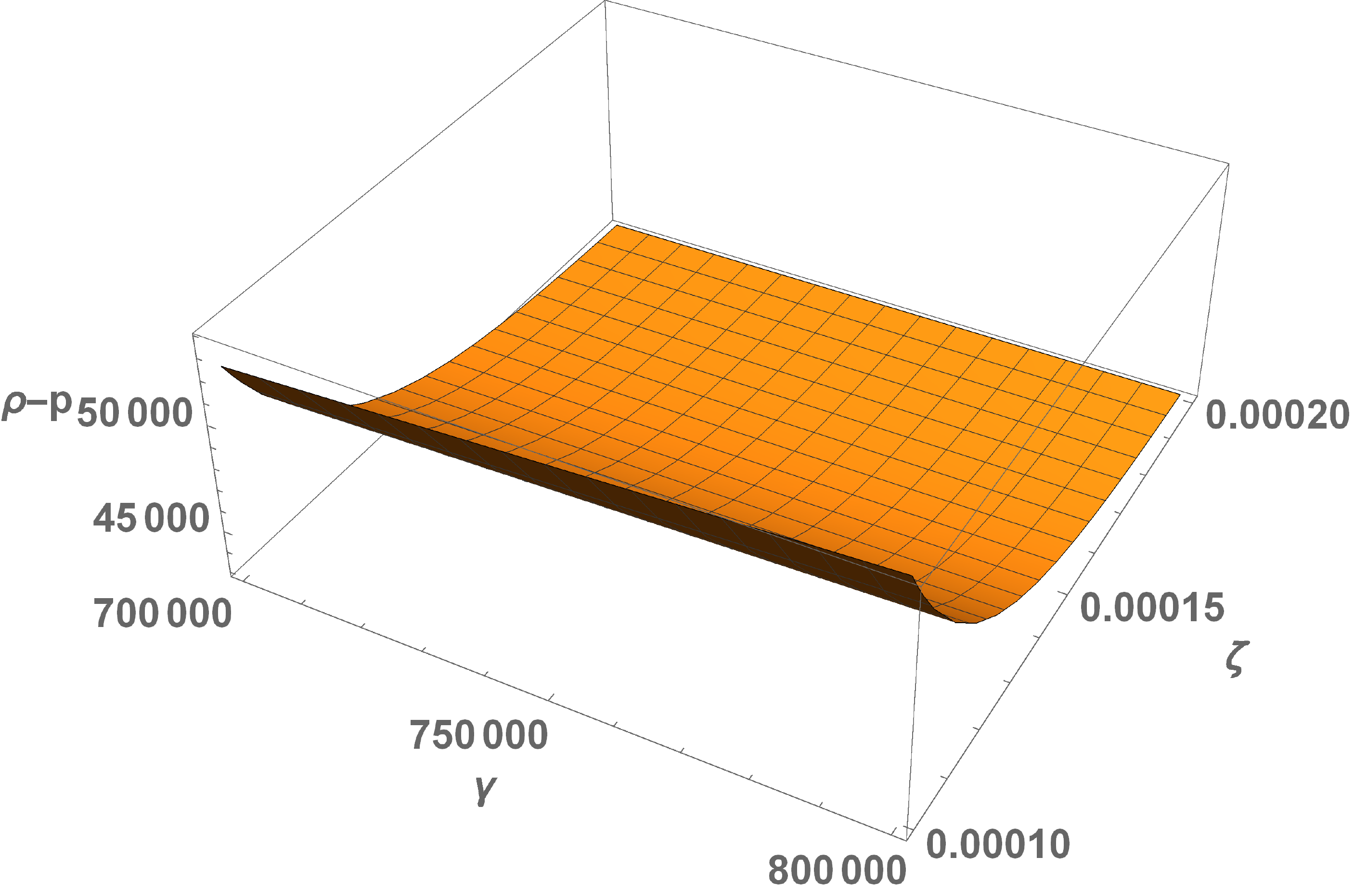}
         \caption{$\rho$ - p.}
         \label{fig:DEC case2 model2}
     \end{subfigure}
     \hfill
     \begin{subfigure}[b]{0.45\textwidth}
         \centering
         \includegraphics[width=\textwidth]{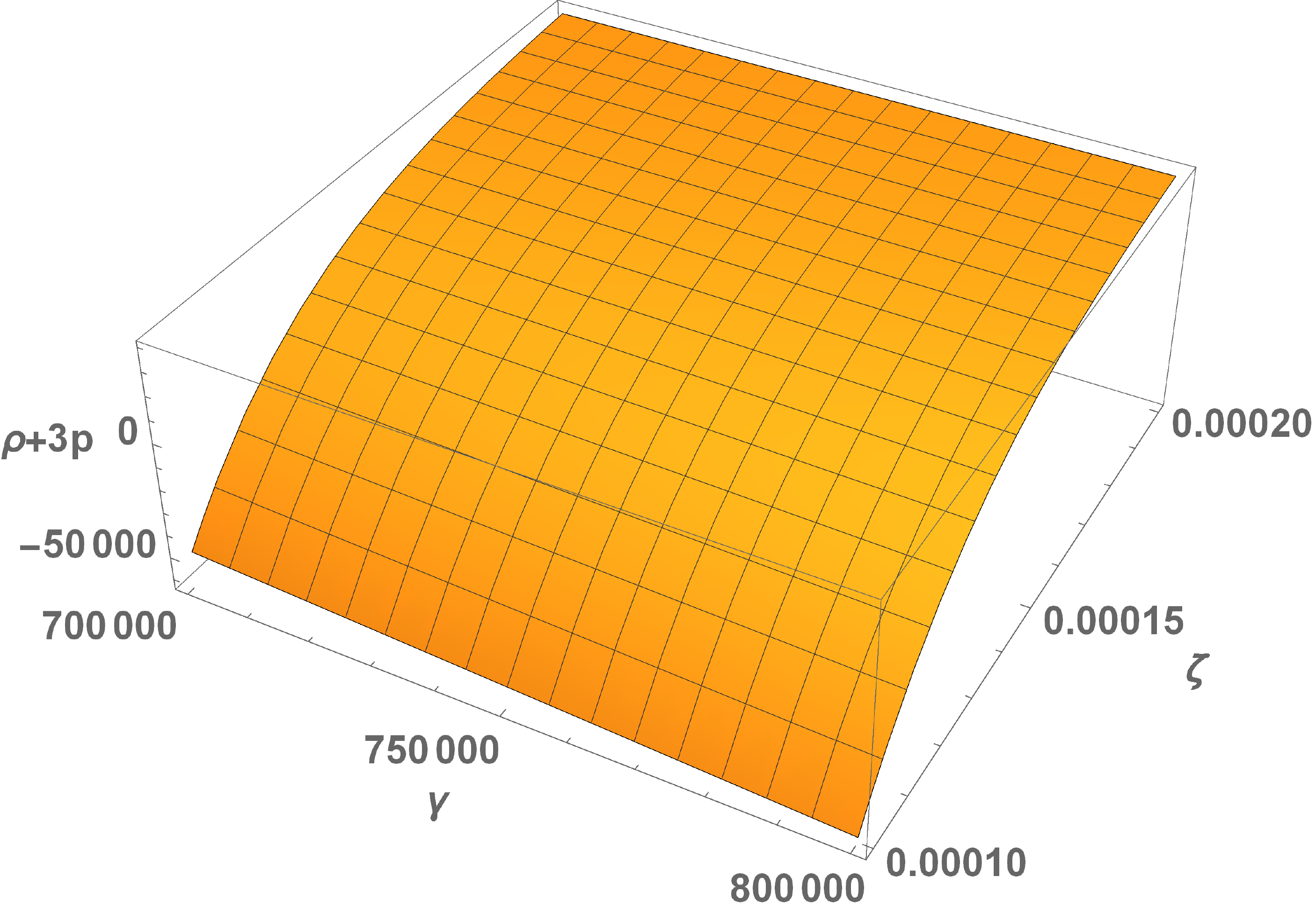}
         \caption{$\rho$ + 3 p.}
         \label{fig:SEC case2 model2}
     \end{subfigure}
        \caption{The behavior of energy conditions considering $F(R)= \gamma Exp(\zeta/R) -R$ for Case II with $ 7 \times 10^{5} \leq \gamma \leq 8 \times 10^{5}$  and $0.0001 \leq \zeta \leq 0.0002$.}
        \label{fig:EC case2 model2}
\end{figure}

Here we can observe from Figs. \ref{fig:NEC case2 model2} \& \ref{fig:DEC case2 model2} that WEC, DEC  are validating their respective conditions with $ 7 \times 10^{5} \leq \gamma \leq 8 \times 10^{5}$  and $0.0001 \leq \zeta \leq 0.0002$. We have considered the bounds of parameters large due to higher powers of $H$. On the other hand, SEC in Fig. \ref{fig:SEC case2 model2} shows negative behavior violating its condition in the same range. Hence, the violation of the SEC supports an accelerated expansion of the universe.

\section{Discussion} \label{sec5}
There is a need to investigate some fundamental questions about the unexplained dark energy, the underlying acceleration of the universe. What is causing this sudden acceleration? Is it due to the addition of a dark energy component such as scalar field or quintessence, or is it due to a modification in Einstein's General Theory of Relativity? In this work, we attempt to investigate the accelerated expansion in the realm of modified $F(R)$ gravity. We assume the  functional form of $F(R)= R+ \alpha  log(\beta R)$ where $\alpha$ and $\beta$ are constants and $F(R)= \gamma Exp(\zeta/R) -R$ with $\gamma$ and $\zeta$ are constants. The spacetime is assumed to be pseudo-symmetric, and a detailed analysis of it has been carried out, providing some insight into the behavior of density and equation of state parameters. The density shows positive behavior, whereas the EoS parameter shows a negative behavior, lying in the quintessence phase in model I whereas a phantom regime in model II, indicating the acceleration in the universe.
Further energy conditions are investigated using the current estimated values of deceleration, jerk, and snap parameters. We obtain the bounds of model parameters $\alpha$, $\beta$, $\gamma$ and $\zeta$ showing the viability of the considered models. However, it is seen that all the energy conditions NEC, WEC, and DEC are satisfying their derived conditions while SEC shows a violation supporting the accelerated expansion of the universe in the Model I. Whereas in the Model II, DEC is met but WEC and SEC are not obeyed, depicting the behavior similar to the scalar tensor gravity models and the accelerating phase, respectively.


\section*{Acknowledgments}  SA acknowledges CSIR, New Delhi, India for JRF.


\end{document}